\documentclass[sigconf]{acmart}
\AtBeginDocument{%
  }

\copyrightyear{2026}
\acmYear{2026}
\setcopyright{cc}
\setcctype{by}
\acmConference[CHI '26]{Proceedings of the 2026 CHI Conference on Human Factors in Computing Systems}{April 13--17, 2026}{Barcelona, Spain}
\acmBooktitle{Proceedings of the 2026 CHI Conference on Human Factors in Computing Systems (CHI '26), April 13--17, 2026, Barcelona, Spain}
\acmPrice{}
\acmDOI{10.1145/3772318.3793420}
\acmISBN{979-8-4007-2278-3/2026/04}

\usepackage{enumitem}
\usepackage{xspace}
\usepackage{xparse}
\usepackage{float}
\usepackage{subfig}
\usepackage{multirow}
\usepackage{gensymb}
\usepackage[ruled,linesnumbered,commentsnumbered]{algorithm2e}
\usepackage[nameinlink]{cleveref}

\newcommand{\revised}[1]{{\color{blue}#1}}
\renewcommand{\revised}[1]{#1} 

\newif\ifanonymized
\anonymizedtrue

\newcommand{\ModeSwitchingCondition}{\textsc{Mode Switching}\xspace}
\newcommand{\OrientationCondition}{\textsc{Orientation Control}\xspace}
\newcommand{\TargetDepthCondition}{\textsc{Target Depth}\xspace}
\newcommand{\ModeButton}{\textsc{Button}\xspace}
\newcommand{\ModeFlip}{\textsc{Flip}\xspace}
\newcommand{\OrientationRoll}{\textsc{StylusRoll}\xspace}
\newcommand{\OrientationPenPoint}{\textsc{StylusPoint}\xspace}
\newcommand{\OrientationGazePoint}{\textsc{GazePoint}\xspace}
\newcommand{\SwitchInTime}{\textsc{Switch In Time}\xspace}
\newcommand{\PositioningTime}{\textsc{Positioning Time}\xspace}
\newcommand{\OrientationTime}{\textsc{Orientation Time}\xspace}
\newcommand{\SwitchOutTime}{\textsc{Switch Out Time}\xspace}
\newcommand{\TrialCompletionTime}{\textsc{Task Completion Time}\xspace}
\newcommand{\PositioningError}{\textsc{Positioning Error}\xspace}
\newcommand{\OrientationError}{\textsc{Orientation Error}\xspace}

\begin{document}

\title{StylusPort: Investigating Teleportation using Stylus in VR}

\author{Yang Liu}
\orcid{0009-0005-9839-0864}
\affiliation{%
    \institution{Aarhus University}
    \city{Aarhus}
    \country{Denmark}
}
\email{yang.liu@cs.au.dk}

\author{Qiushi Zhou}
\orcid{0000-0001-6880-6594}
\affiliation{%
    \institution{The Hong Kong University of Science and Technology (Guangzhou)}
    \city{Guangzhou}
    \country{China}
}
\email{qiushizhou@hkust-gz.edu.cn}

\author{Mathias N. Lystbæk}
\orcid{0000-0001-6624-3732}
\affiliation{%
    \institution{Aarhus University}
    \city{Aarhus}
    \country{Denmark}
}
\email{mathiasl@cs.au.dk}

\author{Aidan Kehoe}
\orcid{0000-0002-6581-3833}
\affiliation{%
    \institution{Logitech}
    \city{Cork}
    \country{Ireland}
}
\email{akehoe@logitech.com}

\author{Mario Gutierrez}
\orcid{0009-0001-6709-0982}
\affiliation{%
    \institution{Logitech Europe S.A.}
    \city{Lausanne}
    \country{Switzerland}
}
\email{mgutierrez1@logitech.com}

\author{Hans Gellersen}
\orcid{0000-0003-2233-2121}
\affiliation{%
  \institution{Lancaster University}
  \city{Lancaster}
  \country{U.K.}
}
\affiliation{%
  \institution{Aarhus University}
  \city{Aarhus}
  \country{Denmark}
}
\email{h.gellersen@lancaster.ac.uk}

\author{Ken Pfeuffer}
\orcid{0000-0002-5870-1120}
\affiliation{%
  \institution{Aarhus University}
  \city{Aarhus}
  \country{Denmark}
 }
\email{ken@cs.au.dk}

\renewcommand{\shortauthors}{Liu et al.}

\begin{abstract}

With a stylus, users can both sweep sketches across models and pinpoint locations with precision. Building on this dual capability, we explore how teleportation can be integrated into stylus interaction without disrupting the flow of common stylus usage. We introduce two key ideas: flipping the stylus as an intuitive mode switch between drawing and teleportation, and using gaze to set orientation while the stylus handles positioning. In a user study that features a teleport-and-\revised{orient} task, we evaluate six teleportation techniques, covering two mode-switching methods (\ModeButton and \ModeFlip) and three orientation approaches (\OrientationRoll, \OrientationPenPoint, and \OrientationGazePoint). The results offer new insights into the relative merits and limitations of each technique. Our work contributes to knowledge about teleportation in VR and fills the gap in seamlessly integrating teleportation with stylus use in 3D.


\end{abstract}


  
\begin{teaserfigure}
  \includegraphics[width=\textwidth]{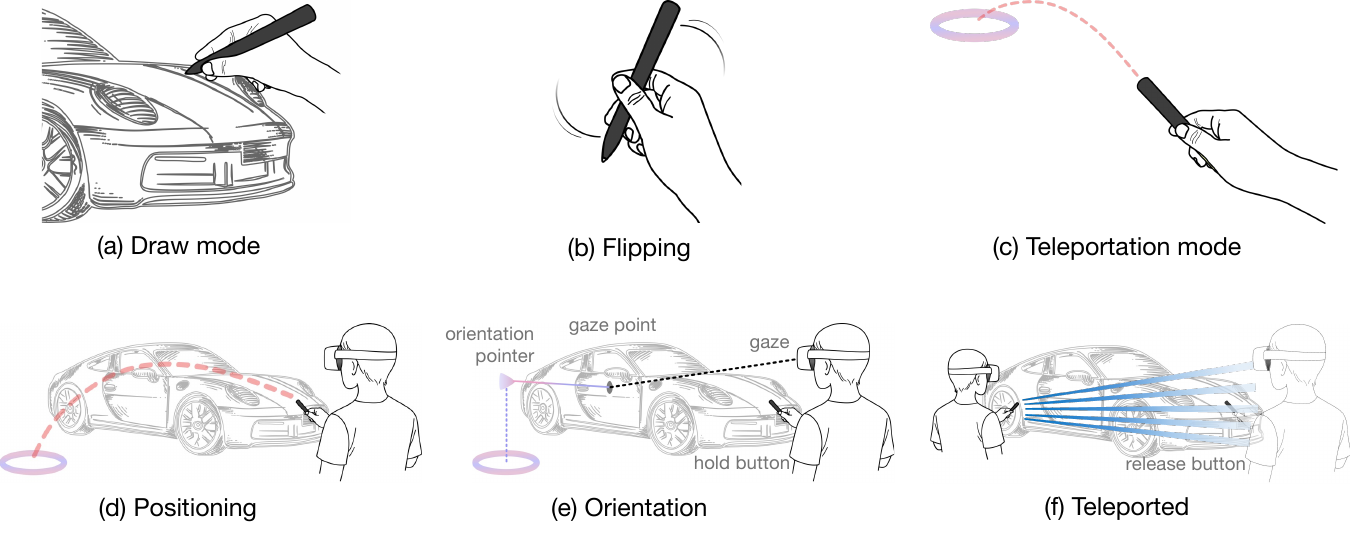}
  \caption{We investigate and propose new ways to teleport with a stylus. For example, by default users sketch (a), but at any time they can flip the stylus to switch to teleport mode (b–c). Users select a target position via raycasting (d). While holding the button, they can look in the direction they want to face after teleporting (e). Releasing the button executes the teleport (f). }
  \label{fig:teaser}
\end{teaserfigure}

\maketitle

\section{Introduction}
Designers can engage in \revised{Virtual Reality} (VR) to work with 2D and 3D content in ways that go beyond traditional \revised{desktop} tools--moving around objects, seeing them from different perspectives, editing, drawing, and refining precise models. A key part of this process is navigation. While users can physically walk \revised{in VR}, they also perform \textit{Point \& Teleport} \cite{Bozgeyikli16, Funk19, mori2023point} to instantly change \revised{position} and \revised{orientation}, \revised{particularly when physical movement is constrained, for example, when sitting on a chair or using a tethered headset}. \revised{Orientation control is especially important when working with large-scale models or navigating large spaces, which is extensively supported in current VR design applications, such as Arkio, EngageXR, and VXRLabs \cite{Arkio, ENGAGE, VictoryXR}. In a typical workflow of such, a designer may teleport inside a large model to focus on a specific component for sketching and editing. They may also need to maneuver within a room-scale scene containing multiple objects (e.g., furniture) and quickly orient themselves around different items of interest.}

This can be performed with a controller that includes a clear division of labour between object manipulation and teleportation capabilities through multiple buttons and joysticks \cite{Matviienko22, Meta25NLocomotion}. With a stylus, however, it is more challenging to map the same number of functions to fewer inputs. A stylus usually has few buttons, while at least one is reserved to activate manipulation in 3D space. A direct one-to-one mapping of controller functionality to a stylus would only make it work for the simplest applications \cite{Kehoe2025Logitech}. And while it is possible to extend a stylus with additional buttons, widgets, menus, or even non-dominant hand input, they take up space and can be error-prone \cite{vogel10,Cami18}, rather than supporting the use of the stylus in a simple or efficient way. 

In this work, we investigate StylusPort: teleportation techniques with stylus-based interaction in VR. Building on prior research on Point \& Teleport techniques using \revised{hand- or }controller-pointing with parabolic rays \cite{Bozgeyikli16, Funk19, mori2023point}, we explore this approach with \revised{VR} stylus\revised{es}, and aim to address two research questions:

\textbf{RQ1 - Mode switching}: How can users switch modes between draw mode and teleportation mode? Previous work in 3D stylus input indicates high potential to leverage implicit methods (e.g., gesture, posture, and grip) \cite{Cami18}. We explore a \textit{flip gesture} and compare it to a \textit{button-based mode switch} as a baseline. 

\textbf{RQ2 - Orientation control}: How can users specify teleportation orientation with a stylus? We investigate three techniques: \OrientationRoll as a means to adjust orientation, extending prior controller \cite{Funk19, mori2023point} and 2D stylus \cite{Bi08} techniques; a \textit{dual-pointer approach} \OrientationPenPoint, where users specify both position and orientation through pointing \cite{mori2023point}; and a \textit{gaze-enabled approach} \OrientationGazePoint, where users look at the point they wish to face after teleporting, inspired by prior work on gaze-directed flying \cite{Templeman99,mine95, Kang2024RayHand} and multimodal gaze + pen work \cite{Pfeuffer2015Shift, Qiu25, wagner25pen}.

To address these questions, we conducted a user study ($N=18$) with two main independent variables: \ModeSwitchingCondition (\ModeButton, \ModeFlip) and \OrientationCondition (\OrientationRoll, \OrientationPenPoint, and \OrientationGazePoint). In each task trial, participants (1) switched to teleportation mode, (2) specified a teleport position and confirmed with a button press, (3) specified orientation until button release, and (4) switched back to draw mode to complete the trial by drawing a stroke. The target was a blackboard representing a typical drawing canvas, presented at \revised{five} possible rotations and two distances, which is inspired by common use \revised{cases} in commercial education applications~\cite{ENGAGE,VictoryXR}.
Teleport positions were constrained to the ground plane, while both the ground and object surfaces could be used to specify orientation.

Our results show that \ModeFlip provided faster mode switching and higher pointing accuracy, thanks to the palm grip, while maintaining an overall \revised{task} completion time comparable to \ModeButton. Regarding orientation control, \revised{we primarily see differences in time efficiency, where} \OrientationRoll proved to be the slowest method, 
\revised{while \OrientationPenPoint was found to be faster for orientation than \OrientationGazePoint.}

Overall, our contributions are: (1) novel teleportation techniques featuring stylus-flip mode switching and gaze-based orientation specification; (2) empirical insights from a user study that delineate performance and user experience trade-offs of these techniques.


\section{Related Work}
\revised{Our work sits at the intersection of VR teleportation, stylus-based interaction, and gaze-based locomotion.
}

\subsection{Teleportation}

Early teleportation \revised{prototypes} started from viewport control in 2D interfaces for 3D applications. For instance, \textit{Navidget} and \textit{Immersive Navidget} explored novel 2D and 3D viewport control approaches when navigating a virtual environment~\cite{Knödel2008Navidget,Hachet2008Navidget}. These early works indicated that positioning and orientation are two fundamental components of teleportation that need to be addressed separately.
This orientation approach has inspired later works. \textit{Anchored Jumping} specifies a \revised{ground point} to define the facing direction \revised{and} then \revised{selects} the \revised{teleport destination} ~\cite{Bimberg2021Virtual}. \revised{As well}, \textit{SkyPort} evaluated linear and parabolic pointing and different transition types\revised{--}instant, interpolated, and continuous\revised{-- and} concluded that linear aiming \revised{with} instant transitions \revised{offers} high efficiency and accuracy without increasing sickness~\cite{Matviienko22}. \citeauthor{Weissker2023Gaining} explored teleportation to \revised{mid-air} 3D positions by specifying ground position and height \revised{either} simultaneously or separately. They reported that simultaneous control \revised{improves} accuracy \revised{but leads to} a longer time than controlling position and height separately~\cite{Weissker2023Gaining}. 

2D teleportation--selecting \revised{a} ground position using a parabol\revised{ic ray} from \revised{the} controller or \revised{the user's} hand--is a common \revised{locomotion method across prevalent} devices and applications. For example, Meta Quest devices introduce 2D teleportation as a \revised{primary} locomotion type implemented \revised{for} both controller and hand microgesture~\cite{Meta25NLocomotion}. These techniques have been iteratively developed and evaluated in HCI \revised{research}. \citeauthor{Bozgeyikli16} proposed \textit{Point \& Teleport} and compared it with 
walk-in-place and joystick locomotion. They also proposed and evaluated an orientation component that let users set their facing \revised{direction} \revised{by} rolling the hand. Whereas Point \& Teleport \revised{was found} fun and user-friendly, the orientation component \revised{was considered unintuitive and difficult to control}~\cite{Bozgeyikli16}. Similarly, a study on spatial cognition found that teleportation is less error-prone \revised{without} orientation \cite{Cherep2020Spatial}.

\revised{For pointing methods of 2D teleportation, \citeauthor{Rupp2025HowFar} found that linear rays make long-distance teleportation more difficult than parabolic rays \cite{Rupp2025HowFar}, while other studies reported that linear rays are faster and more intuitive in certain contexts \cite{Sermarini21Kinematic, Narbayev25}. Despite that, parabolic rays are now the teleportation standard.} \citeauthor{Funk19} \revised{investigated parabolic-ray Point \& Teleport with several orientation control techniques:} \textit{AngleSelect} \revised{(touchpad-based)}, \textit{Curved Teleport} that visualizes \revised{an adjustable curved} trajectory, and \textit{HPCurved} that combines a parabola with an adjustable curvature. They argued that \revised{while} these three techniques with orientation \revised{control} were slower than \revised{those} without, they \revised{reduced post-teleport orientation} correction~\cite{Funk19}. \citeauthor{mori2023point} evaluated \revised{four} orientation approaches for Point \& Teleport \revised{inspired by} ``natural turning'', \revised{including} head-turning, integrated touchpad \revised{control}, wrist\revised{-based} orientation, and \revised{post-positioning} pointing for orientation. They concluded that \revised{while} orientation is useful--especially in extreme conditions--no \revised{single} consensus \revised{of approach is both} intuitive \revised{and low}-effort~\cite{mori2023point}. \citeauthor{Muller2023UndoPort} explored undoing teleportation \revised{with} separate position and orientation components, finding that combining \revised{both} significantly improves \revised{usability} \cite{Muller2023UndoPort}. Finally, most \revised{commercial} VR applications \revised{implement orientation as a separate step from positioning}~\cite{ENGAGE,VictoryXR,GravitySketch}. In this work, we address the gap in integrated positioning and orientation with a \revised{stylus.} 



\subsection{Stylus-based Interaction}

Whereas mid-air stylus input was explored \revised{in} early desktop VR systems \cite{Deering1995HoloSketch}, the large virtual spaces afforded by modern VR present new opportunities and challenges \revised{for using} 3D stylus input in navigation while performing primary \revised{tasks} such as sketching and manipulation \cite{Keefe2007drawing, Romat21Flashpen}. Early work addressed this by reducing the need to move \revised{physically and} enabling stylus input  at \revised{a distance through physical surfaces}.  For instance, \citeauthor{Arora2017Experimental} \revised{demonstrated} the viability of using a physical drawing surface to support free-form mid-air sketching in VR \cite{Arora2017Experimental, Arora2018SymbiosisSketch}. Other works explored multitouch gestures, such as \textit{VRSketchIn}, \revised{which} investigated a design space of stylus and tablet interaction for 3D sketching in VR that combines unconstrained 3D mid-air with constrained 2D surface-based sketching \cite{Drey20VRSketchIn}. However, 3D sketching \revised{without} a physical surface is still a major use case, \revised{as} eviden\revised{ced by} leading \revised{VR} applications such as ShapesXR, GravitySketch, and VXRLabs~\cite{ShapesXR,GravitySketch,VictoryXR}, broadly used in 3D product design. Other works have explored mid-air 3D interaction with \revised{distant} objects \revised{using a stylus, such as pointing and manipulation}. \citeauthor{Chen22Investigating} found that VR controllers and styluses yield significantly higher precision \revised{than bare hands} in a 3D target-tracing task, suggesting that while styluses benefit precise 3D drawing, target selection \revised{may be better handled by} hand or another modality due to \revised{stylus-induced} fatigue~\cite{Chen22Investigating}. 

\revised{Integrating stylus input with other modalities opens new interaction opportunities. \citeauthor{Matulic2023PenTouchMidairCH} explored bimanual pen-and-touch interactions in VR, showing how asymmetric pen-hand coordination supports manipulation and navigation \cite{Matulic2023PenTouchMidairCH}. \citeauthor{wagner25pen} investigated combining gaze and stylus input for selecting and translating shape points in 3D modeling, finding that gaze-assisted dragging reduced task time and manual effort with slightly increased errors \cite{wagner25pen}. We extend the prior multimodal interaction work by a focus on teleportation.}

Similar to the use of physical pens, \revised{grip postures also affect comfort and performance} for different tasks, such as coarse and precise drawing. \citeauthor{Batmaz20Precision} found that a ``precision grip'' (typical pen grip) significantly improves accuracy in VR \cite{Batmaz20Precision}. \citeauthor{Li20Grip} similarly found that a rear-end ``tripod'' precision grip allows the largest range of motion. They specified that while forward-and-downward pointing is easy with a precision grip, forward-and-upward pointing is easier with a palm grip (like holding a wand), which also induces less fatigue during prolonged use~\cite{Li20Grip}. 

Previous work has \revised{explored} enriching \revised{stylus} affordances by \revised{using} different grip postures for \revised{distinct} functions and modes. \citeauthor{Cami18} evaluated common variations of precision grips and \revised{demonstrated} using \revised{unoccupied} fingers for alternative input modes, such as touch on a tablet \cite{Cami18}. \citeauthor{Song_2011_Grips} proposed \textit{Multi-Touch Pen}, \revised{which} enables \revised{input mode} switching through different tapping gestures on a stylus afforded \revised{by} different grip postures \cite{Song_2011_Grips}. \citeauthor{Cai25HPIPainting} presented \textit{HPIPainting} that detects different grips to contextualize \revised{gesture} recognition \revised{for} triggering commands in VR painting \cite{Cai25HPIPainting}. \citeauthor{Li2005Mode} specifically investigated switching between inking and gesturing mode when using a stylus. They explored approaches such as changing \revised{stylus tip} pressure, holding the stylus still, and pressing a button using the non-dominant hand. They found trade-offs between approaches \revised{with} no clear winner \revised{in} speed and accuracy \cite{Li2005Mode}. Other works evaluated alternative approaches, including non-dominant hand gestures and inferring from \revised{stylus} trajectory for mode switching between inking and other input (e.g., selection), suggesting that mode switching needs more thoughtful design than simply adding gestures or using implicit information \cite{Saund03Stylus,Smith20Evaluating}. 

In this work, inspired by previous work on input mode switching for VR \cite{Surale2019Mode} and other platforms \cite{Pfeuffer20}, we explore a novel approach of switching between drawing and teleportation modes using a stylus flip action that transitions between precision grips and palm grips. \revised{This exploits} the ease of forward-and-upward pointing using the palm grip, \revised{which} suits parabola-based Point \& Teleport techniques~\cite{Li20Grip}.

\subsection{Gaze-based Locomotion}

Gaze input has been envisioned in some of the earliest explorations of VR locomotion. In his seminal work \textit{Virtual environment interaction techniques}, Mine envisioned \textit{gaze-direct\revised{ed} flying}, \revised{where users are transported towards} the direction of their gaze~\cite{mine95}. Similarly, an early exploration of ``walking in place'' proposed \revised{incorporating gaze and head direction}---\revised{essential} natural behaviors during locomotion---\revised{as an integrated} input modality~\cite{Templeman99}. Following a similar idea of using gaze for viewport control, \citeauthor{Lee2024Snap} explored \revised{several gaze-based} viewport control \revised{techniques} combined with head movement, including snapping to gaze-dwell location, using gaze saccades as gain functions to amplify head rotation, and gaze pursuit \revised{for} aligning the viewport with specific targets~\cite{Lee2024Snap}. These works suggest a promise of using gaze as an additional modality for teleportation in VR.

Gaze has been explored as a natural and easy-to-use input modality, \revised{both} implicit \revised{and} explicit, to specify points of interest in interaction tasks \cite{Jacob16,Jacob90}. One particular use case relevant to our work is extending direct input devices like touch, gesture, or pen with additional indirect input modes. Previous works, such as Gaze-Touch \cite{Pfeuffer2014touch}, Gaze-Shifting \cite{Pfeuffer2015Shift}, and Gaze+Pinch \cite{Pfeuffer17pinch}, exploit the natural eye-hand coordination so users can access both direct and gaze-assisted indirect input modes, enabling rapid \revised{access to distant} objects. In the large virtual spaces \revised{of} VR, this  saves time and effort induced by manual pointing while retaining the primary role of the given input modality \cite{wagner2024eye,Lystbaek24}. 


These benefits of gaze input have been explored in recent works on VR teleportation. For instance, \citeauthor{Kim2023Exploration} explored teleportation methods using hand-tracking, eye-tracking, and EEG \revised{signals}. Their proposed techniques used \revised{either} hand or gaze \revised{for} position \revised{aiming}, \revised{and triggered teleportation} using hand gestures or EEG. They found that gaze\revised{-based} positioning was faster and more precise than hand pointing \cite{Kim2023Exploration}. \citeauthor{Lee2024RPG} proposed an approach of specifying \revised{teleportation} orientation by detecting users' gaze towards directions that fall out of the central area of their view, constructing it as an explicit trigger \revised{for orientation} change \cite{Lee2024RPG}. In this work, we explore the use of gaze for orientation specification in VR teleportation as an explicit pointer \revised{that activates after stylus-based positioning}.


\section{Design of  Stylus-based Teleportation}

We \revised{describe} the design \revised{of techniques for switching to teleportation mode and specifying orientation with a stylus.}

\subsection{Mode Switching}
\revised{We assume typical stylus-based interaction 
where most of the available buttons are dedicated to default operations (e.g., sketching, pointing, object manipulation), and incorporate a mode-switching button for users to enter teleportation mode. In this mode, users point with the stylus to specify position and press-hold a button to specify orientation, and release to accomplish the teleport. }

\subsubsection{\ModeButton} 
\revised{Our rationale for the first mode-switching technique is as follows. 
We ruled out assigning teleportation to a simple button press because users need to preview the parabolic teleport ray before performing a teleport, and continuously displaying this ray would interfere with primary actions such as sketching. 
Showing the ray by pressing and holding the button is also not viable--the button-hold action is already needed for orientation control. 
%
Instead, it is reasonable for the secondary button, namely the button not dedicated to primary actions, to act as a mode selector. Our first teleportation method, \ModeButton, adopts this approach: pressing the button switches from the primary function into teleportation mode; pressing it again returns to the previous mode. In principle, additional modes could be added to provide more functionality beyond teleportation. Within teleportation mode, the ray visualization gives clear feedback that the user is in the correct mode.
}

\revised{\subsubsection{\ModeFlip} For teleportation without relying on a button}, we exploit \revised{stylus grip semantics}. Prior work \revised{shows} that different stylus grips afford \revised{distinct functions and can support mode switching}~\cite{Cami18, Li20Grip, Song_2011_Grips, Cai25HPIPainting}. Building on this, we employ the natural distinction between \revised{drawing} with the stylus tip and pointing with the stylus tail: \revised{teleportation is active whenever the stylus is flipped}. This \ModeFlip method \revised{provides a low-cost, button-free mechanism that leverages finger dexterity and familiar pen usage. A flip action} is illustrated in \autoref{fig:teaser}a–c.
\revised{\ModeFlip is detected by monitoring the direction of the stylus (from tail to tip) relative to the user's forward view. A flip is valid when the stylus points more than 120$\degree$ away from the camera's forward vector. The angular threshold was chosen through in-house testing to minimize unintentional flip events.}
%
%
\\\\
\revised{For both \ModeButton and \ModeFlip, a} parabolic ray \revised{cast from the stylus previews the teleport destination.}
The \revised{ray} trajectory is defined by the origin (stylus tip or tail, depending on the grip), initial direction (tail-to-tip or tip-to-tail), velocity (10 $m/s$), gravity (9.81 $m/s^2$), and a maximum fall time (1.5 s). These yield a maximum \revised{distance} of about 11.15 m at an optimal pitch angle of 42$\degree$. When the parabola hits the ground, a solid circular cursor appears and marks the teleport destination. \revised{Users press the primary function button--which acts as the teleport button in teleportation mode--to confirm the destination, and release it to execute the teleport.}

To support teleporting behind objects, we allow \revised{developers to} configure whether specific objects block or permit the parabol\revised{ic ray}. For example, walls block teleportation, while a blackboard may allow the ray to pass. This provides a shortcut for crossing obstacles without detouring. \revised{As shown in \autoref{fig:TeleportBehind},} when the parabol\revised{ic ray} penetrates an object, that object temporarily becomes semi-transparent \revised{so users can clearly see} the destination. 

\subsection{Orientation Control}\label{sec:OrientationControl}

\revised{We design techniques where} a quick click \revised{of the teleport button} triggers a teleport without \revised{changing the user's orientation}, while a long press enables orientation control. This mirrors a two-stage input design, as also seen in \citeauthor{mori2023point}'s \textit{P2T}, which differentiates between half- and full-press actions \cite{mori2023point}. We set the threshold between a short click and a hold to $200$ ms, slightly above the $150$ ms average click duration \revised{observed in our in-house testing}.
In the following, we illustrate the design of three orientation methods based on this principle, \revised{namely \OrientationRoll, \OrientationPenPoint, and \OrientationGazePoint}.

\begin{figure*}
    \centering
    \includegraphics[width=1\linewidth]{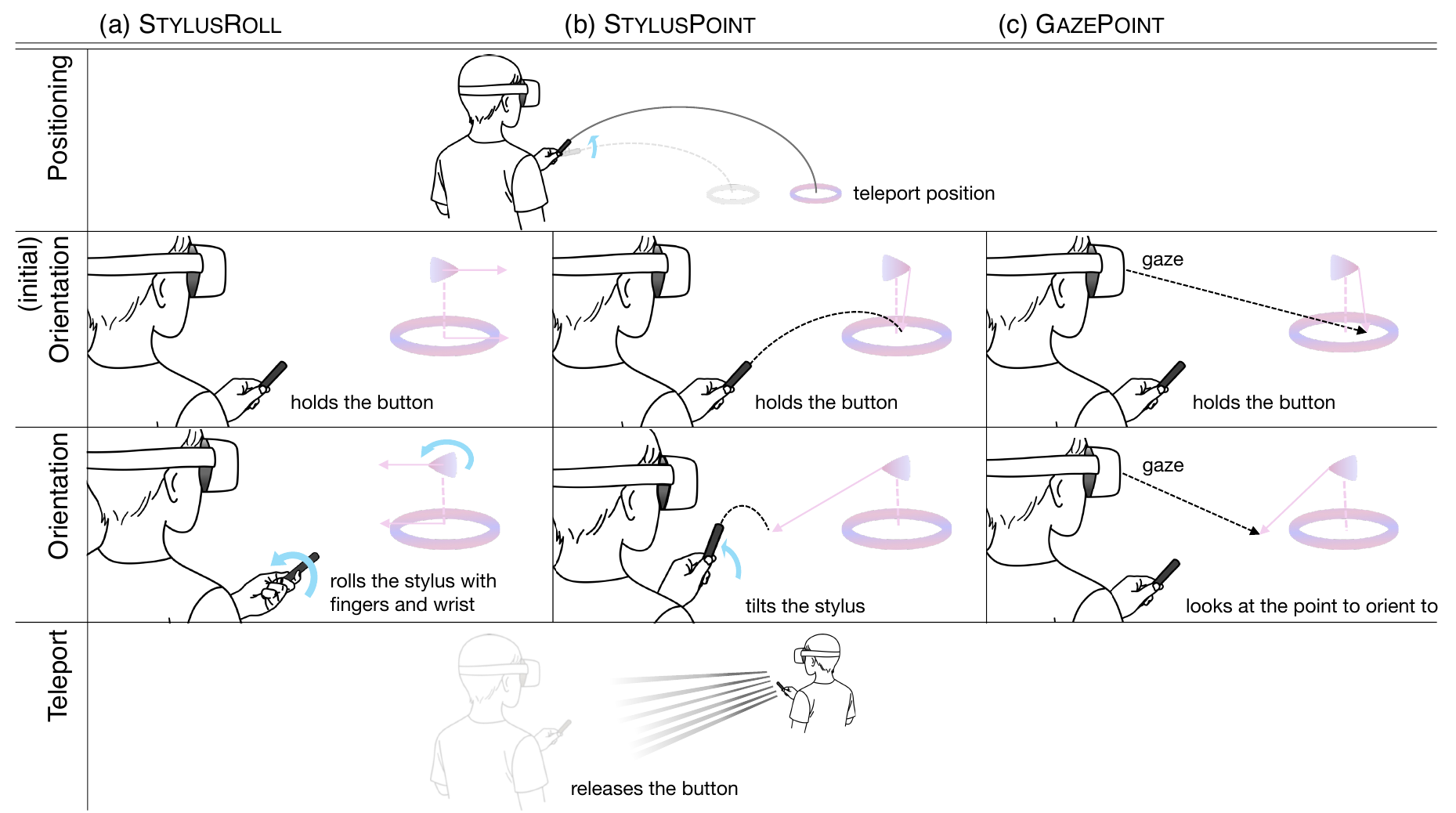}
    \caption{Interaction steps for teleportation with three orientation controls (example with mode switch by stylus flip): \OrientationRoll, \OrientationPenPoint, and \OrientationGazePoint. After the mode switch, the user first confirms a teleport position by pressing and holding the teleport button. Then, they specify the orientation by either (a) rolling the stylus, (b) pointing the stylus toward the desired facing direction, or (c) directing their gaze toward the desired facing direction. Releasing the button triggers the teleport. \revised{Note that the blue arrows are for illustrative purposes and are not displayed to users.}}
    \label{fig:InteractionSteps}
\end{figure*}



\subsubsection{\OrientationRoll}
\OrientationRoll\xspace  \revised{ builds on prior work ~\cite{Bozgeyikli16, Funk19, mori2023point} that maps stylus roll to relative changes in} orientation \revised{(cf. \autoref{fig:InteractionSteps}a), leveraging the} unused rotational axis during pointing. 
\revised{When the user holds the button, the initial orientation is set to} the user's head direction. \revised{Rolling the stylus then rotates this orientation either left or right according to the stylus's rotational direction.}
\revised{The technique supports flexible rolling not only through wrist rotation, but also through in-hand finger movement.}
%
To 
\revised{facilitate large orientation angles, we align with \citeauthor{mori2023point}~\cite{mori2023point} and} apply a $1.5\times$ scaling factor: \revised{$1\degree$ rolling maps to $1.5\degree$ orientation change. This was determined through pilot testing to balance rolling efficiency and accuracy}.
%
\revised{The visual feedback  is illustrated in \autoref{fig:InteractionSteps}a}. A 3D arrow is displayed above the teleport destination \revised{to indicate orientation}, \revised{appearing} 20 cm below the user's \revised{eye level}. Also, two \revised{1 m} horizontal \revised{lines} extend from the center and the arrow in the direction.

\subsubsection{\OrientationPenPoint} 
\OrientationPenPoint introduces an orientation cursor. 
Once the teleport button is held, the position cursor \revised{on the ground} becomes fixed. \revised{Then}, a second parabolic ray\revised{--}with the same parameters as the \revised{positioning parabola--}is cast to define an \revised{orientation} cursor \revised{(cf. \autoref{fig:InteractionSteps}b)}. \revised{The cursor appears at the intersection point with any surface, including the ground and objects}.
The orientation is defined by the \revised{horizontal} vector from the destination to the orientation cursor. 
%
   \revised{Upon button release, users are teleported to the destination and oriented toward the orientation cursor.}

\begin{figure}
    \centering
    \includegraphics[width=\linewidth]{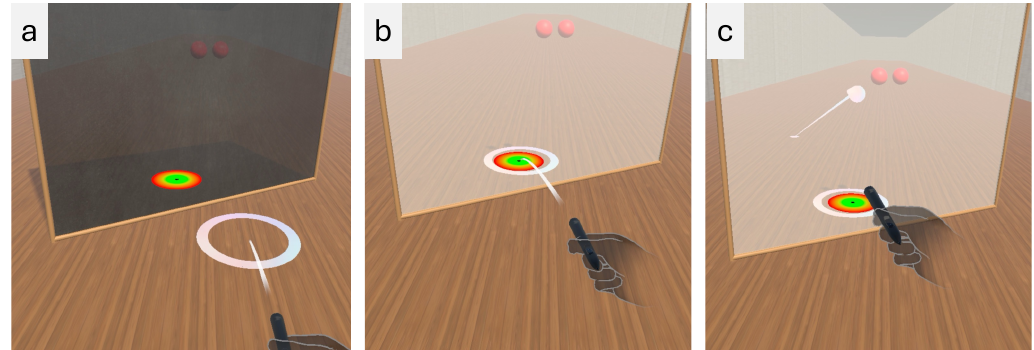}
    \caption{When teleporting behind an object, the user positions the cursor so that the ray intersects the object. The object then becomes semi-transparent, allowing the orientation ray to pass through. }
    \vspace{-20pt}
    \label{fig:TeleportBehind}
\end{figure}

\revised{As illustrated in \autoref{fig:TeleportBehind}}, when a user aims to teleport behind an object, the object becomes semi-transparent upon being penetrated by the \revised{positioning} ray, thereby \revised{allowing the orientation ray to also pass through.}
\revised{This thereby} ensur\revised{es} a consistent behavior between position and orientation pointing. 

\revised{The visual feedback of \OrientationPenPoint is presented as} a 3D arrow above the position cursor pointing in the horizontal direction toward the orientation cursor, which is rendered as a solid circle and connected to the arrow with a line \revised{(cf.~\autoref{fig:InteractionSteps}b)}. 
\revised{Since the orientation cursor can be cast on any surface, displaying it along with a connection line facilitates cursor placement and orientation perception.}
To reduce clutter, the orientation parabolic ray is not rendered. 

\subsubsection{\OrientationGazePoint}

\OrientationGazePoint extends \OrientationPenPoint with multimodal input by using gaze for orientation, \revised{instead of a stylus-based pointing parabolic ray}. This method leverages the natural behavior of gaze, as users typically look at \revised{where} they wish to face before teleport\revised{ing}.
After a position is \revised{selected via stylus pointing and} confirmed by \revised{pressing} the button, \revised{users specify their post-teleport orientation using gaze during button holding. A gaze ray is continuously cast and can intersect with any surface, including the ground and objects, similar to the orientation ray in \OrientationPenPoint. The orientation is defined by the horizontal vector from the destination position to the gaze cursor. Upon button release, users are teleported to the selected position and oriented. }
%
\OrientationGazePoint's \revised{visuals} are identical to \OrientationPenPoint, \revised{illustrated in \autoref{fig:InteractionSteps}c}.  

\section{User Study}

We evaluate the mode switching and teleportation orientation \revised{techniques using} a teleport-and-\revised{orient} task. 






\subsection{Study Design} 
We designed a within-subject experiment with two independent variables: \ModeSwitchingCondition (\ModeButton, \ModeFlip) and \OrientationCondition  (\OrientationRoll, \OrientationPenPoint, \OrientationGazePoint). To reduce order effects, we counterbalanced all conditions using a Latin Square. Participants first experienced the \OrientationCondition conditions in a counterbalanced order, and within each \OrientationCondition condition, they completed both \ModeSwitchingCondition conditions in counterbalanced order.
Each \ModeSwitchingCondition $\times$ \OrientationCondition combination consisted of \revised{10} trials (2 target depths $\times$ \revised{5} target orientations). We repeated each trial set twice, resulting in a total of $2$ target depths $\times\ \revised{5}$ target orientations $\times\ 2$  mode switching $\times\ 3$ orientation controls $\times\ 2$ repetitions $= \revised{120}$ trials per participant.

\begin{figure*}[h]
    \centering
    \includegraphics[width=1\linewidth]{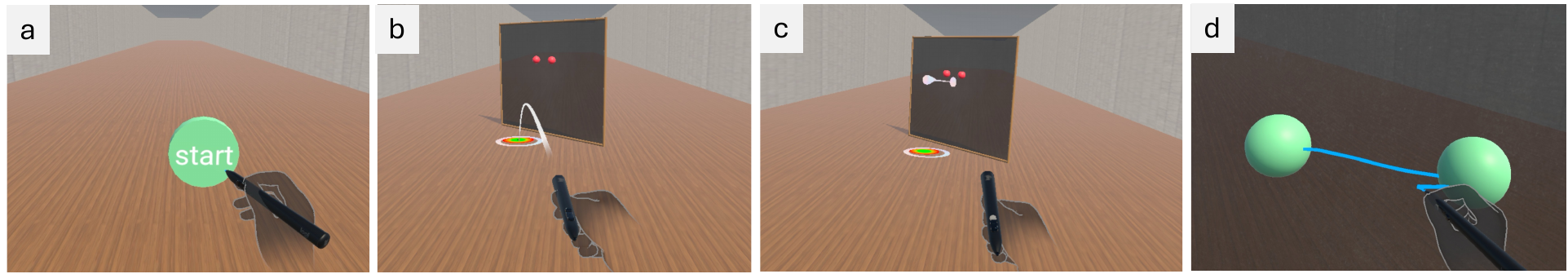}
    \caption{Study \revised{example with} \ModeFlip\!+\OrientationPenPoint:  (a) The \revised{user} taps the button with the stylus to start, (b) they select a \revised{teleport} destination, (c) they specify orientation \revised{while pressing the button}, (d) \revised{release to} teleport and draw a line to finish.}
    \label{fig:StudyEnv}
\end{figure*}
 
\subsection{Task}
\begin{figure}[t]
    \centering
    \includegraphics[width=\linewidth]{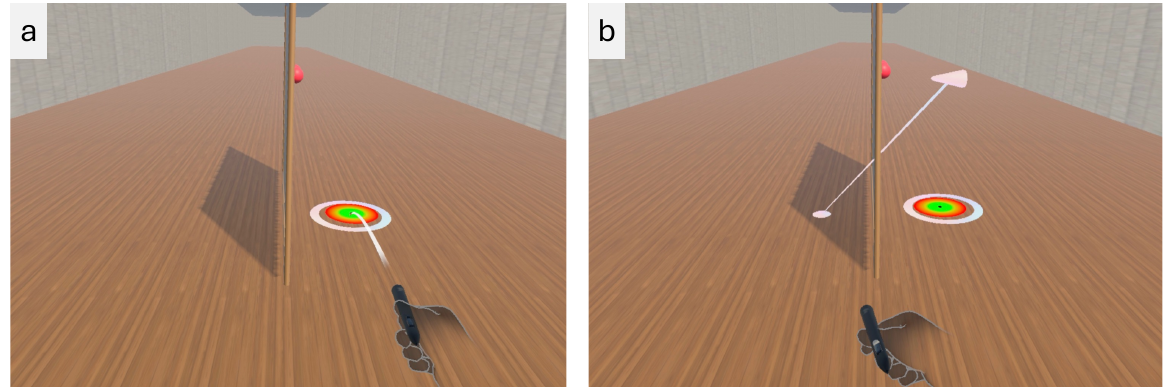}
    \caption{When teleporting to face a blackboard presented from its side, it can be difficult to cast the orientation point onto its thin edge. A practical strategy is to instead place the orientation point on the ground behind the blackboard. }
    \label{fig:TeleportSide}
\end{figure}

\revised{We designed the task in a VR blackboard drawing scenario where stationary, seated participants teleport.}
Participants are immersed in a long corridor ($70m \times 8m$) with wooden flooring, walls, and ceiling. They press a start button \revised{with the stylus} to begin a trial. This reveals a transparent blackboard ($88\%$ alpha, $2m \times 1.5m$) which appears at two \revised{possible} distances ($3m$ and $6m$) and \revised{five}  rotation angles ($45^\circ, -45^\circ$, $90^\circ$, $-90^\circ$, $180^\circ$).
%
Two red spheres on the blackboard's front face are positioned $1m$ above the floor and $30 cm$ apart from each other. A gradient circle \revised{with a black center} marks the teleport \revised{target} $50cm$ in front of the blackboard  \revised{(cf. \autoref{fig:TeleportBehind})}.
\revised{For a trial, p}articipants \revised{1) switch mode} from draw to teleportation, \revised{2)} direct the positioning parabola to the target circle, \revised{3) press} the teleport button to activate orientation control, \revised{4)} orient towards the midpoint between the two spheres, \revised{5)} release the teleport button to teleport, \revised{6)}  switch back to draw mode and connect both spheres in one stroke (successful trial) or draw in mid-air (failed). \revised{After 1.5s, the next trial begins.}
\autoref{fig:StudyEnv} \revised{shows a full trial}, \autoref{fig:TeleportBehind} \revised{shows a $45\degree$ trial}, and  \autoref{fig:TeleportSide} \revised{shows a $-90\degree$ trial}. 


\subsection{Apparatus and Implementation} 
\begin{figure}
    \centering
    \includegraphics[width=0.85\linewidth]{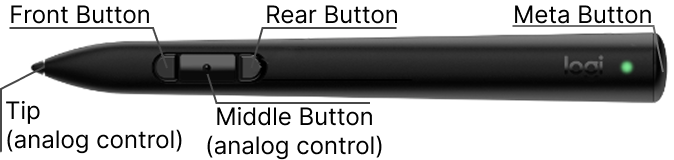}
    \caption{\revised{Logitech MX Ink stylus controls. Figure adapted with permission from \cite{Kehoe2025Logitech}.}}
    \label{fig:MXInkControls}
\end{figure}

We implemented with Unity on the Meta Quest 3, which supports the Logitech MX Ink stylus\footnote{https://www.logitech.com/en-us/products/vr/mx-ink.html}. The stylus is 164 mm \revised{long}, $18.2$ mm in diameter, and weighs $29$ g.
\revised{ (cf. \autoref{fig:MXInkControls})}.
Drawing was performed by pressing the front button. With \ModeButton, users switched modes by clicking the rear button \revised{and teleported with} the front button. 
With \ModeFlip, users flipped the stylus to switch modes, and either button teleported. Teleportation was implemented using the Meta XR Interaction SDK (v77) \revised{Teleport Interaction}. 
\revised{The stylus and the user's hand models were rendered in the scene. }

Because Meta Quest 3 does not natively support eye-tracking, we fitted a third-party Neon XR eye tracker from Pupil Labs\footnote{https://pupil-labs.com/products/vr-ar} ($100$ Hz). \revised{Gaze was smoothed with a 10-sample window to reduce noise. }
The eye tracker \revised{is reported to have} an accuracy of around $1.3 - 1.8^\circ$ \revised{in a 2D-screen setup} (outdoors \revised{degrading by} $0.2^\circ - 0.4^\circ$)~\cite{Baumann23NeonTestReport}. \revised{We conducted a separate accuracy test in VR based on \cite{Baumann23NeonTestReport} using 9 targets spaced 20$\degree$ apart from the center at 5 depths (0.5–2.5 m), finding an average accuracy of 2.68$\degree$ (SD = 2.24$\degree$).}
\revised{The Neon XR provides offset correction to adjust global positional offset of the estimated gaze. We therefore presented a single-point target at a distance of $6m$ to participants, adjusting their gaze offset until they reported accurate eye-tracking. }

\subsection{Procedure}
Participants were briefed on the study context and completed consent and demographics forms. They were shown a video \revised{of the techniques}, along with an explanation of the study setup and a demonstration of how to use the stylus.
\revised{For  \OrientationRoll, participants were shown that they could roll the stylus either by using the fingers or their wrist.}
Participants were seated on a static chair  \revised{to fucus on effects of orientation, minimizing effects of physical orientation}. \revised{They} donned the XR headset and \revised{held} the MX Ink stylus in their dominant hand with a precision grip.
Participants were then presented with a training session matching the task design \revised{and received assistance as needed}. 
\revised{With \OrientationGazePoint, they completed gaze offset correction.}
Users completed at least six training trials, \revised{and more if deemed needed by the experimenter}. 
After \revised{each} condition, they were presented with a post-condition NASA-TLX questionnaire. \revised{After all conditions}, the participant was presented with a final post-study questionnaire, consisting of 7-point preference ratings for all conditions along with \revised{a short interview}. The study lasted one hour on average.

\subsection{Evaluation Metrics}
\begin{itemize}
                \item \SwitchInTime: time taken from trial start to switching from draw mode to teleportation mode.
                \item \PositioningTime: time taken to specify a teleportation position after entering teleportation mode.
                \item \OrientationTime: time taken to specify an orientation after specifying the position.
                \item \SwitchOutTime: time taken to switch back from teleportation mode to draw mode after teleporting.
                \item \TrialCompletionTime: time taken to complete a task,  from pressing the start button to switching back to draw mode after teleport. \revised{Drawing time is excluded as not of interest to the research questions.}
    \item \PositioningError: positional offset \revised{(}meters\revised{)} between the teleport and the \revised{target} center \revised{positions}.
    \item \OrientationError: angular offset (degrees) between the \revised{user}'s forward orientation and the \revised{user} direction to the midpoint of the two spheres.
    \item  NASA-TLX~\cite{Hart1988DevelopmentResearch} \revised{in the form} of Raw-TLX questionnaires~\cite{Hart2006Nasa-TaskLater, Bustamante2008MeasurementTLX}.
    \item Technique ratings on a 7-point Likert scale (1 --  least preferred, 7 --  most preferred) and \revised{a brief interview}. 
\end{itemize}

\subsection{Participants}
We recruited 18 participants ($10$ male, $7$ female, and $1$ non-binary) \revised{from} the local university, mainly Master's students in Computer Science. The ages ranged from 18 to 34, all were right-handed, 8 wore glasses, and 1 wore contact lenses. On a 5-point Likert scale, participants rated themselves as having \revised{medium} experience with VR ($M=3.17, SD=1.04$), tablets/smartphones \revised{styluses} ($M=3, SD=0.91$), \revised{and} little experience with VR styluses ($M=1.39, SD=0.78$) and teleportation ($M=1.61, SD=0.78$).

\section{Results}
We \revised{conducted} a three-way (\ModeSwitchingCondition $\times$ \OrientationCondition $\times$ \TargetDepthCondition) \revised{Aligned Rank Transform (}ART\revised{)} \revised{Repeated Measures} ANOVA for data \revised{analysis}, as measures were non-normally distributed after outlier filtering (\revised{Inter-Quartile Range} based on \TrialCompletionTime; $\approx\!\!5.3\%$), followed by Holm-Bonferroni corrected post hoc tests.  
For NASA-TLX and preference ratings, we performed Friedman with post hoc Wilcoxon tests (Holm-Bonferroni corrected).
Statistical significances in \autoref{fig:switch-in-and-positioning-time}-\autoref{fig:orientation-error-and-preference} are shown as \revised{*/**/*** for $p<.05$/$p<.01$/$p<.001$} and error bars show $95\%$ confidence intervals.

\subsection{\revised{Time Measures}}
\begin{figure}
    \centering
    \subfloat[\ModeSwitchingCondition on \SwitchInTime.]{\includegraphics[width=0.28\linewidth]{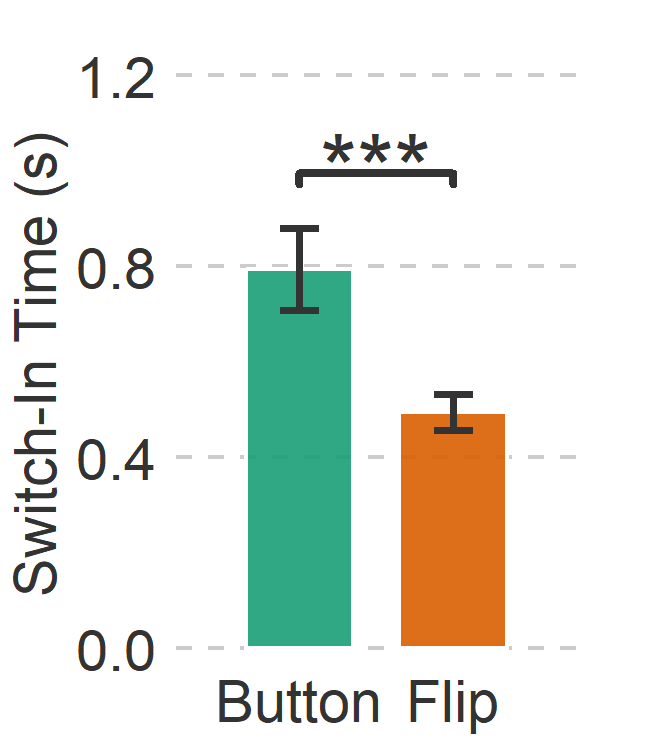}\label{subfig:switch-in-time}}
    \quad
    \subfloat[\ModeSwitchingCondition on \PositioningTime.]{\includegraphics[width=0.28\linewidth]{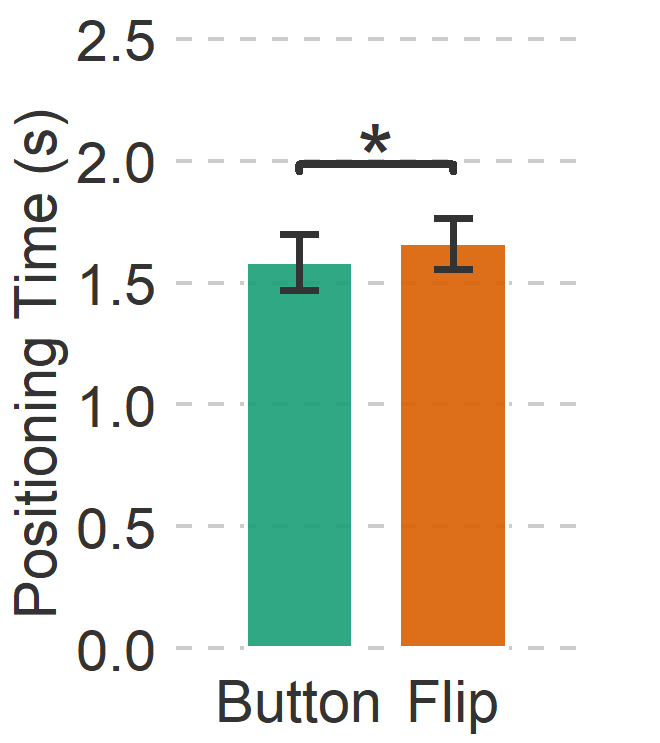}\label{subfig:positioning-time-mode}}
    \quad
    \subfloat[\TargetDepthCondition on \PositioningTime.]{\includegraphics[width=0.28\linewidth]{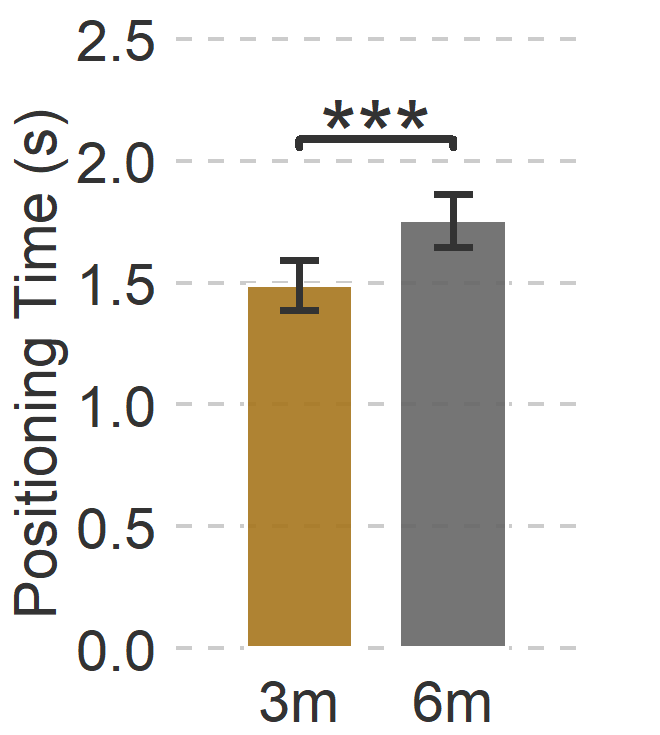}\label{subfig:positioning-time-depth}}
    \caption{Results of on \SwitchInTime (a) and \PositioningTime (b-c). \revised{Notably, main effects show \ModeFlip to switch in faster (a) while being slower at positioning (b) with positioning at 3m to be faster (c)}.}
    \vspace{-30pt}
    \label{fig:switch-in-and-positioning-time}
\end{figure}

\revised{On \SwitchInTime (\autoref{subfig:switch-in-time}),} we found that participants \revised{switched} into teleportation mode \revised{faster when performing the} \ModeFlip (\revised{$F_{1, 187} = 69.25, p < 0.001, \eta^2_p=0.270$}) than \revised{when pressing the} \ModeButton.

\revised{As for \PositioningTime (\autoref{fig:switch-in-and-positioning-time}a-b),} we found that participants were faster with \ModeButton compared to \ModeFlip (\revised{$F_{1, 187} = 6.59, p = 0.011, \eta^2_p=0.034$}).
Participants also took longer to position $6m$ away compared with $3m$ (\revised{$F_{1, 187} = 35.63, p < 0.001, \eta^2_p=0.160$}).

\begin{figure}
    \centering
    \subfloat[\ModeSwitchingCondition.]{\includegraphics[width=0.25\linewidth]{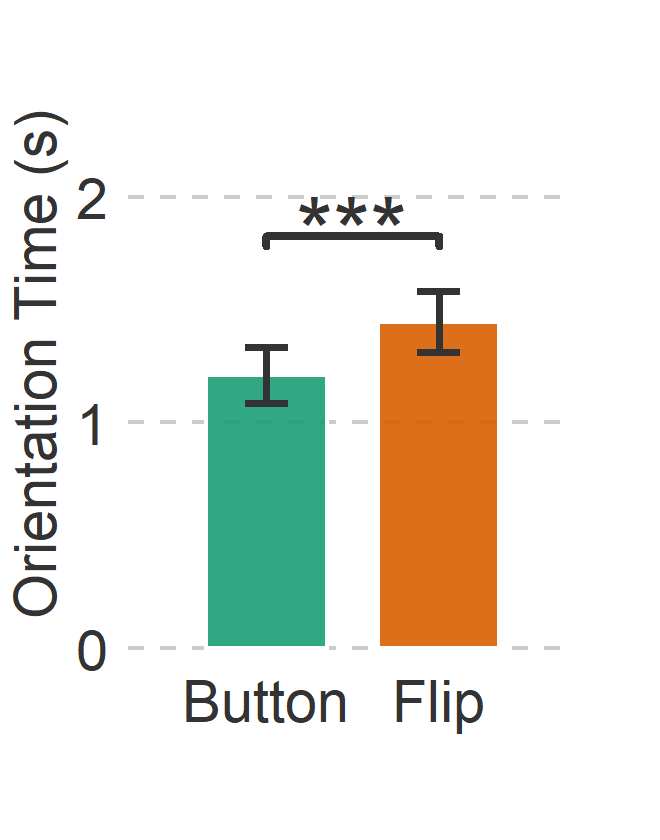}}
    \quad
    \subfloat[\OrientationCondition.]{\includegraphics[width=0.25\linewidth]{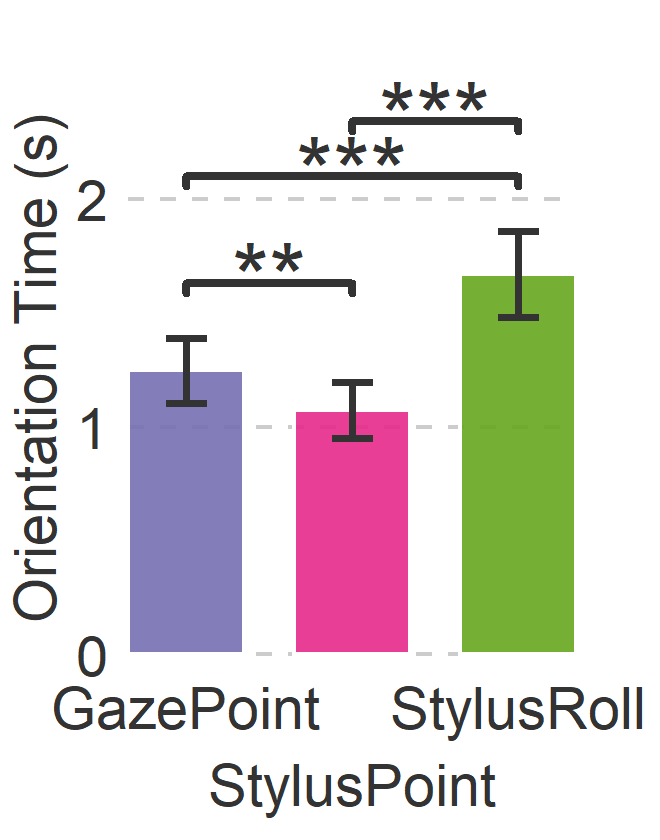}}
    \quad
    \subfloat[\ModeSwitchingCondition $\times$ \OrientationCondition.]{\includegraphics[width=0.4\linewidth]{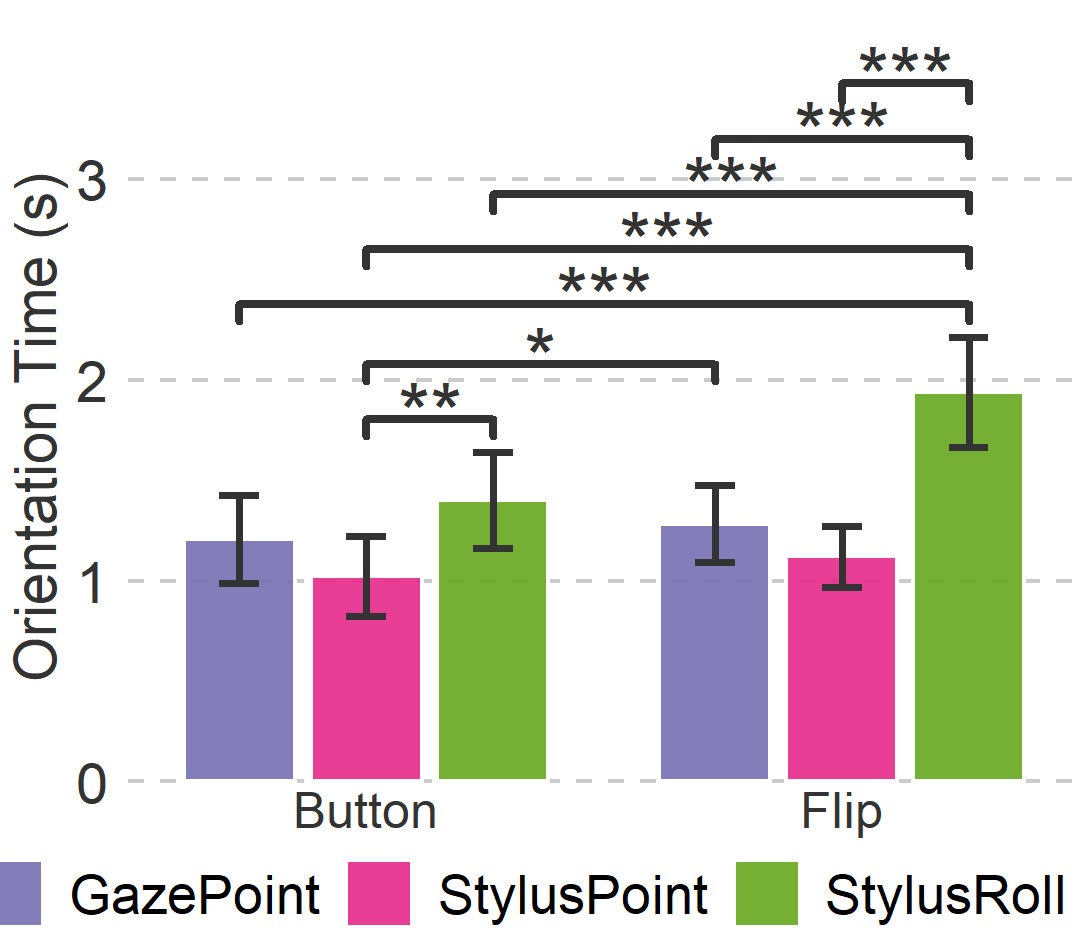}}
    \caption{Results of \OrientationTime. \revised{Notably, main effects show \ModeButton to be faster (a). Meanwhile, \OrientationPenPoint is faster than both \OrientationGazePoint and \OrientationRoll, with \OrientationRoll being the slowest (b)}.}
    \label{fig:orientation-time}
\end{figure}

\revised{Regarding \OrientationTime (\autoref{fig:orientation-time}),} we found that participants \revised{specified the orientation} faster when using \ModeButton (\revised{$F_{1, 187} = 20.54, p < 0.001, \eta^2_p=0.099$}).
\OrientationCondition also had an effect (\revised{$F_{2, 187} = 30.56, p < 0.001, \eta^2_p=0.246$}), as participants were slowest when using \OrientationRoll compared with the others (both $p<0.001$), while participants were faster when using \OrientationPenPoint compared with \OrientationGazePoint (\revised{$p=0.008$}).
\revised{Additionally,} a significant interaction was found between \ModeSwitchingCondition $\times$ \OrientationCondition (\revised{$F_{2, 187} = 4.38, p=0.014, \eta^2_p=0.045$}). Specifically, \OrientationRoll was significantly \revised{more impacted by the grip in \ModeFlip} than \revised{both \OrientationPenPoint and} \OrientationGazePoint (\revised{both $p<0.034$}).

\begin{figure}
    \centering
    \subfloat[\ModeSwitchingCondition on \SwitchOutTime.]{\includegraphics[width=0.24\linewidth]{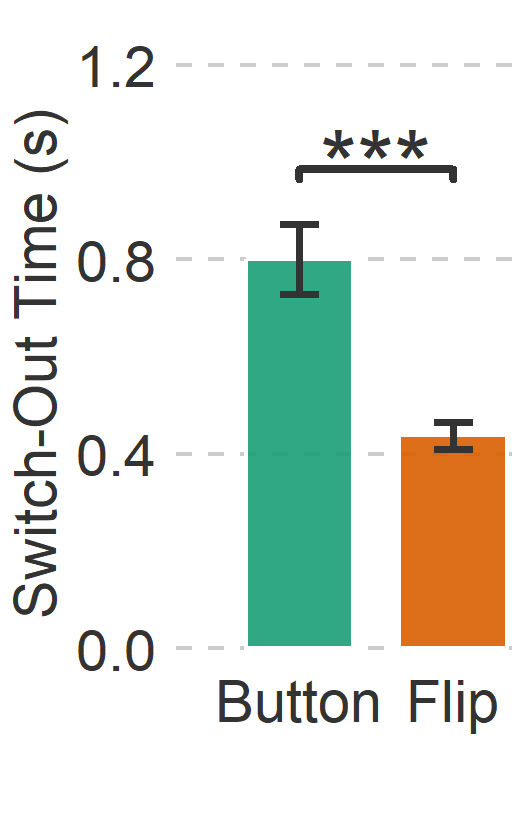}\label{subfig:switch-out-mode}}
    \quad
    \subfloat[\TargetDepthCondition on \SwitchOutTime.]{\includegraphics[width=0.24\linewidth]{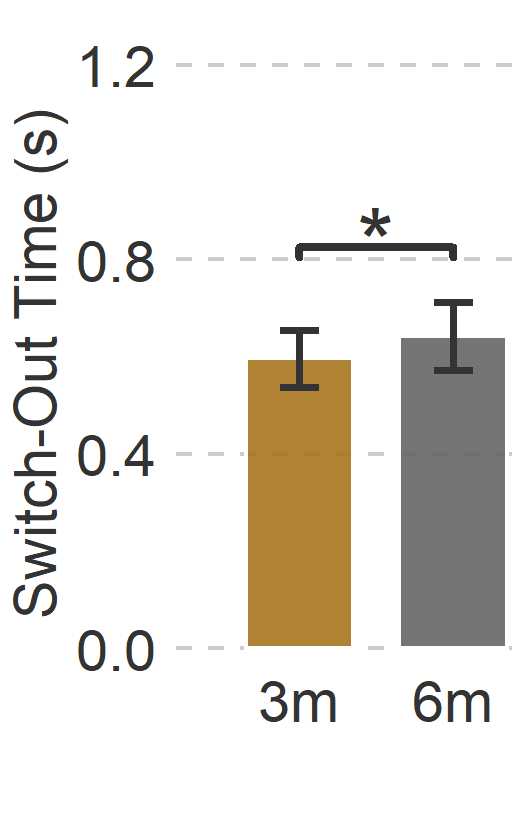}\label{subfig:switch-out-depth}}
    \quad
    \vspace{-12pt}
    \subfloat[\ModeSwitchingCondition $\times$ \TargetDepthCondition on \SwitchOutTime.]{\includegraphics[width=0.425\linewidth]{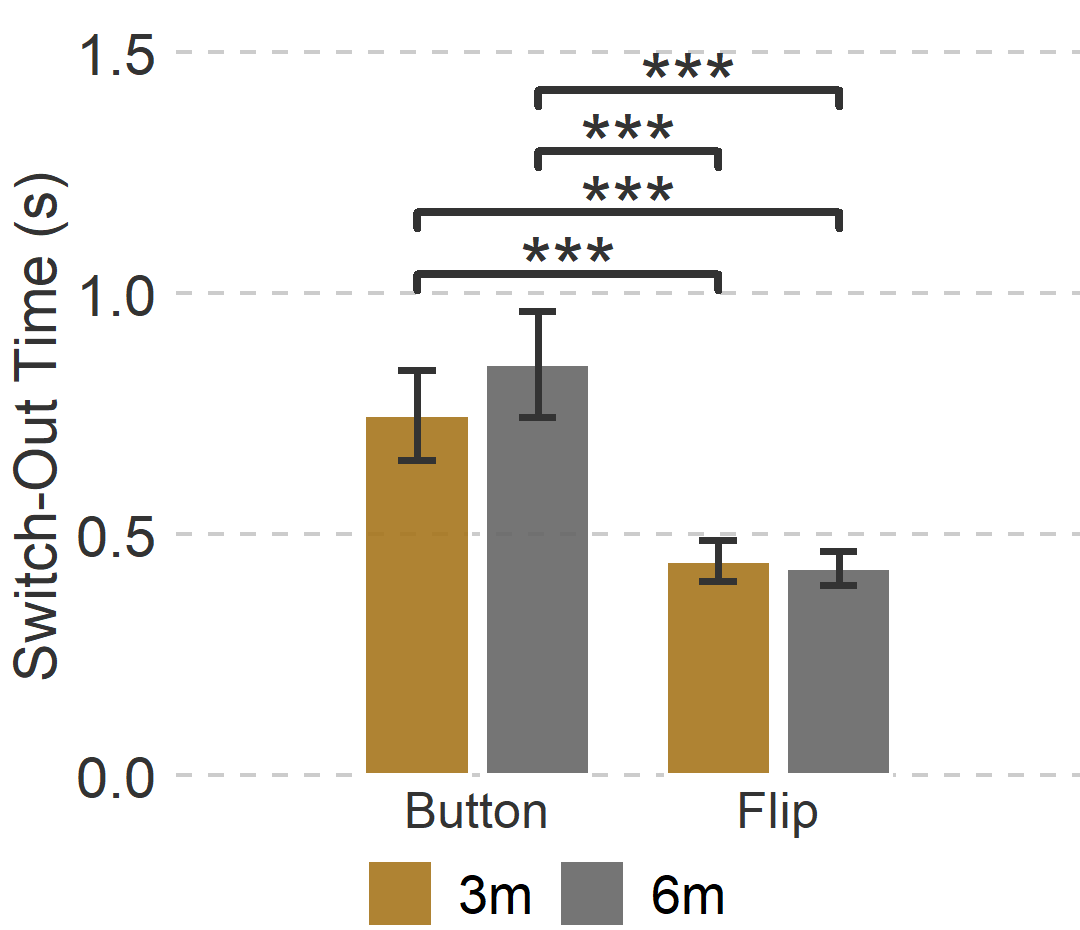}\label{subfig:switch-out-int-mode-depth}}
    \quad
    \subfloat[\OrientationCondition on \TrialCompletionTime.]{\includegraphics[width=0.27\linewidth]{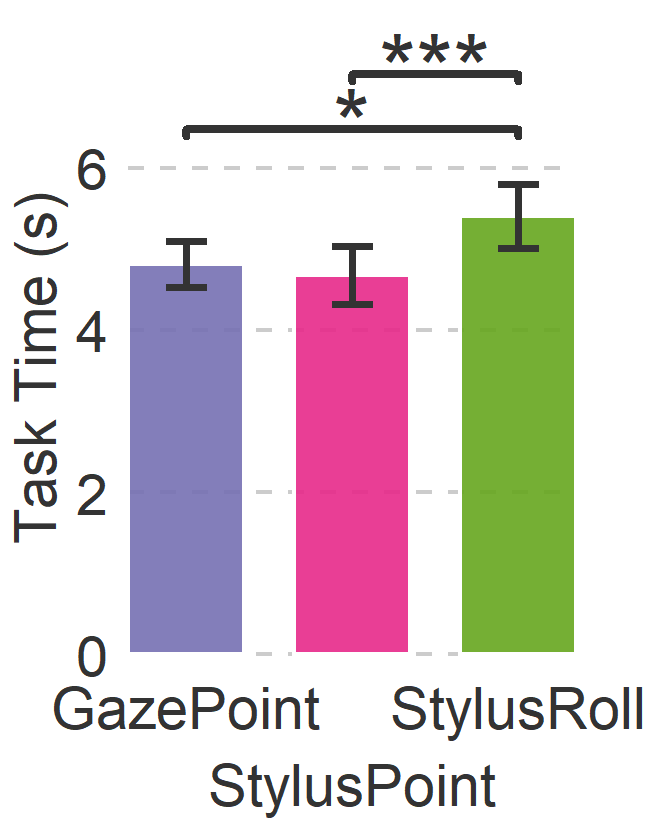}\label{subfig:task-time-ori}}
    \quad
    \subfloat[\TargetDepthCondition on \TrialCompletionTime.]{\includegraphics[width=0.23\linewidth]{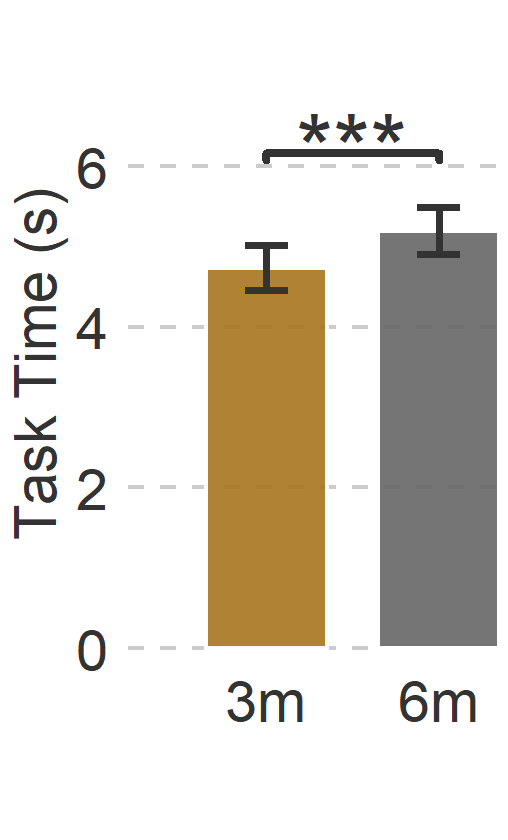}\label{subfig:task-time-depth}}
    \caption{Results of \SwitchOutTime (a-c) and \TrialCompletionTime (d-e). \revised{Notably, main effects show \ModeFlip (a) and 3m (b) to switch out faster, while showing \OrientationRoll to be the slowest overall (a) and teleporting at 3m to be faster overall (c)}.}
    \label{fig:switch-out-time-task-time}
\end{figure}

\revised{When looking at \SwitchOutTime (\autoref{fig:switch-out-time-task-time}a-c),} we found that participants were significantly faster when using \ModeFlip compared with \ModeButton (\revised{$F_{1, 187} = 195.62, p < 0.001, \eta^2_p=0.511$}). 
\TargetDepthCondition also impacted \SwitchOutTime as participants were faster at switching out at the closer distance of $3m$ (\revised{$F_{1, 187} = 4.01, p = 0.047, \eta^2_p=0.021$}).

Further, we found significant interactions between \ModeSwitchingCondition $\times$ \TargetDepthCondition (\revised{$F_{1, 187} = 6.67, p=0.011, \eta^2_p=0.034$}). Specifically, \ModeButton was significantly \revised{more impacted by depth} than \ModeFlip (\revised{$p=0.011$}).

\revised{Finally, on \TrialCompletionTime (\autoref{fig:switch-out-time-task-time}d-e),} we found that \OrientationCondition had a significant effect (\revised{$F_{2, 187} = 9.34, p < 0.001, \eta^2_p=0.091$}), as \OrientationRoll was significantly slower overall than the others (both \revised{$p < 0.018$}).
Participants were also slower overall when teleporting to $6m$ away (\revised{$F_{1, 187} = 14.95, p < 0.001, \eta^2_p=0.074$}).

\subsection{\revised{Error Measures}}
\begin{figure}
    \centering
    \subfloat[\ModeSwitchingCondition.]{\includegraphics[width=0.24\linewidth]{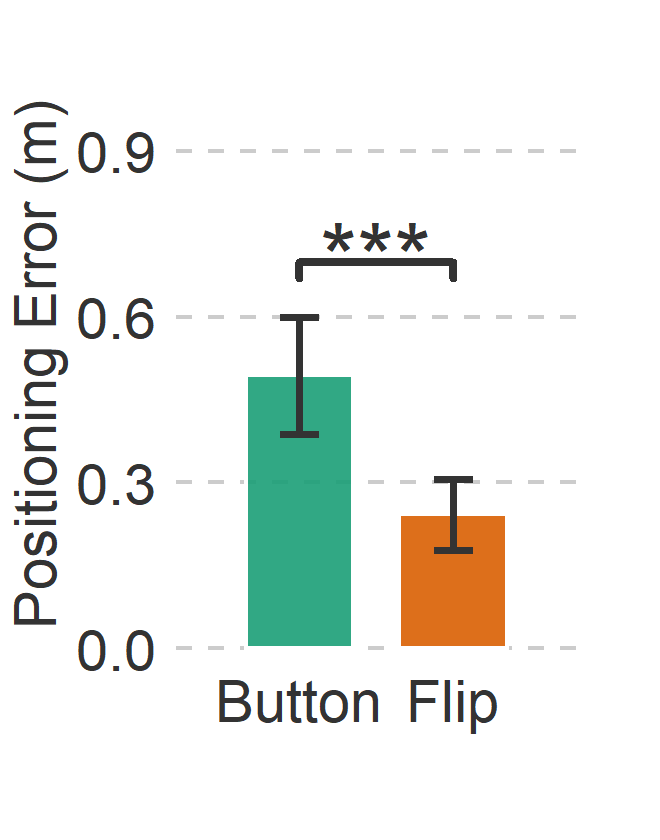}}
    \quad
    \subfloat[\TargetDepthCondition.]{\includegraphics[width=0.24\linewidth]{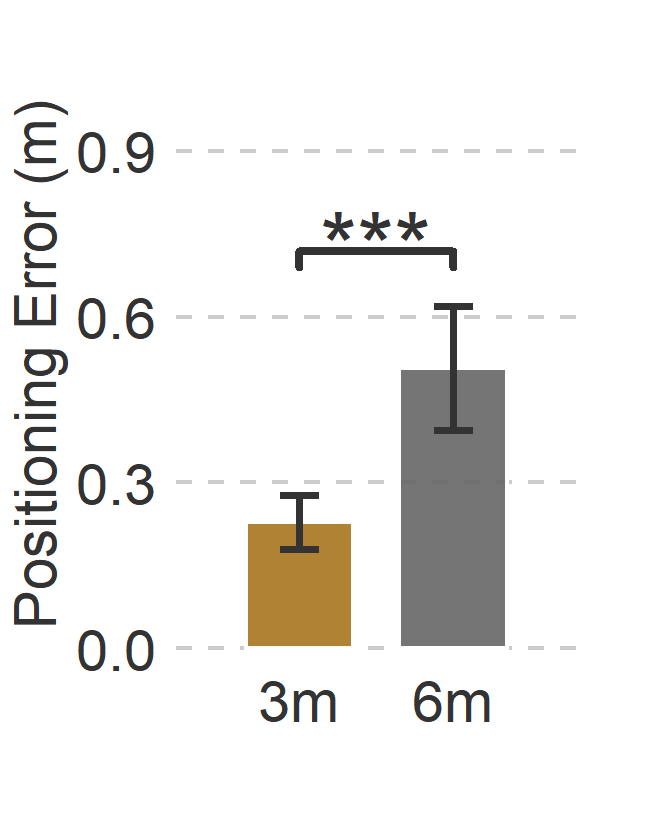}}
    \quad
    \subfloat[\ModeSwitchingCondition $\times$ \OrientationCondition.]{\includegraphics[width=0.4\linewidth]{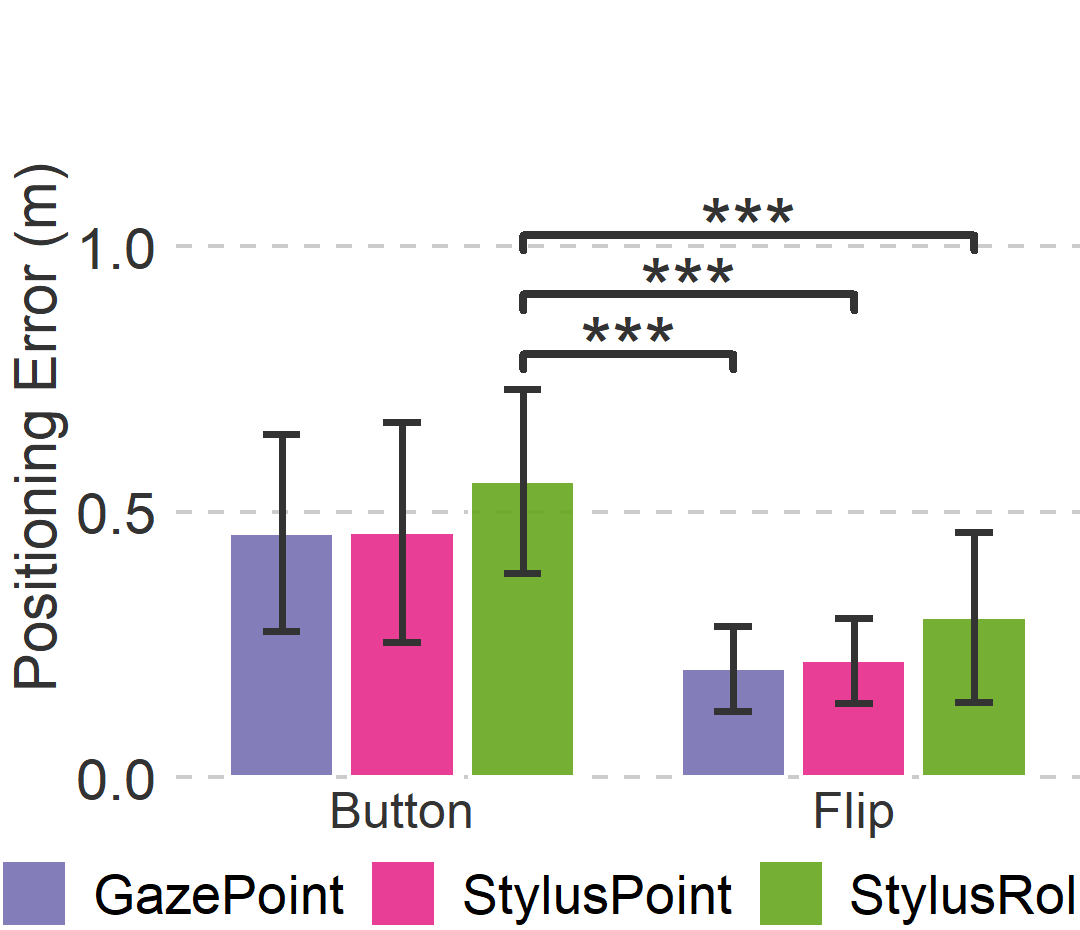}}
    \\
    \subfloat[\ModeSwitchingCondition $\times$ \TargetDepthCondition.]{\includegraphics[width=0.45\linewidth]{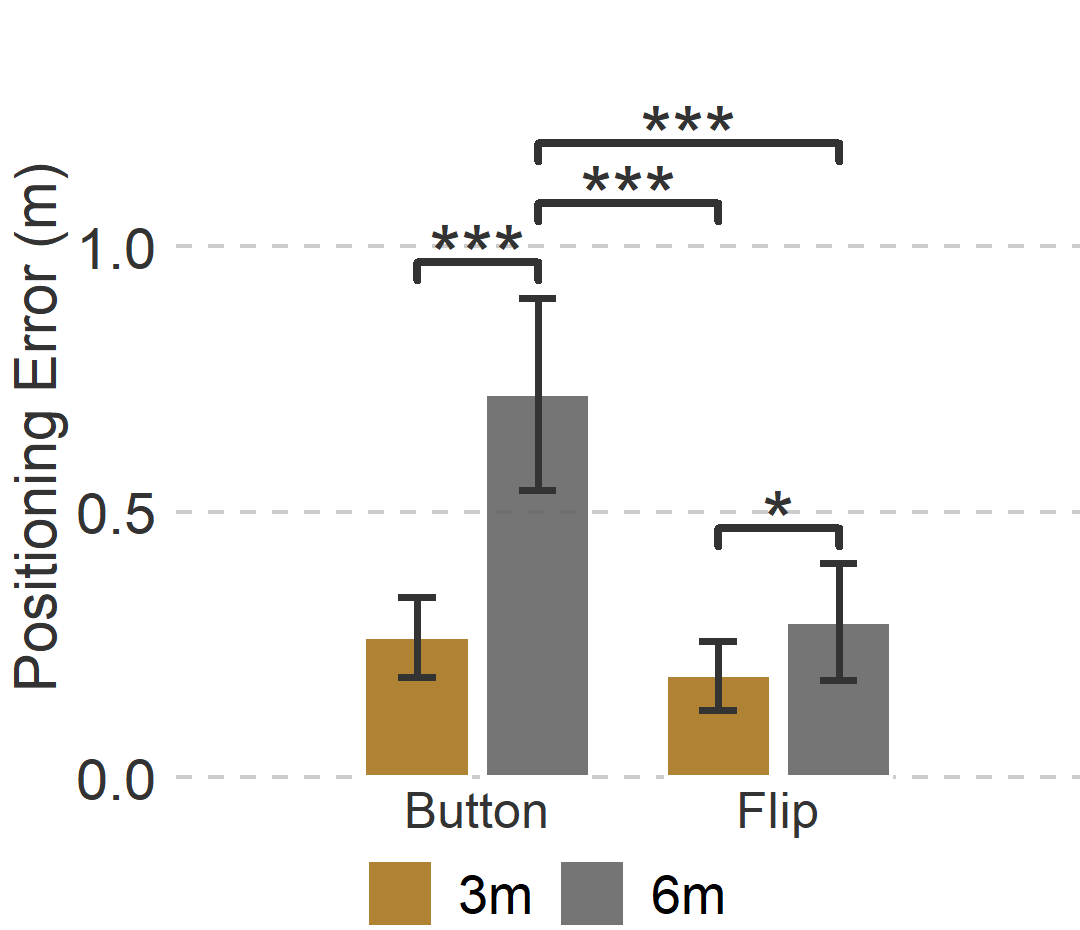}}
    \quad
    \subfloat[\OrientationCondition$\times$ \TargetDepthCondition.]{\includegraphics[width=0.45\linewidth]{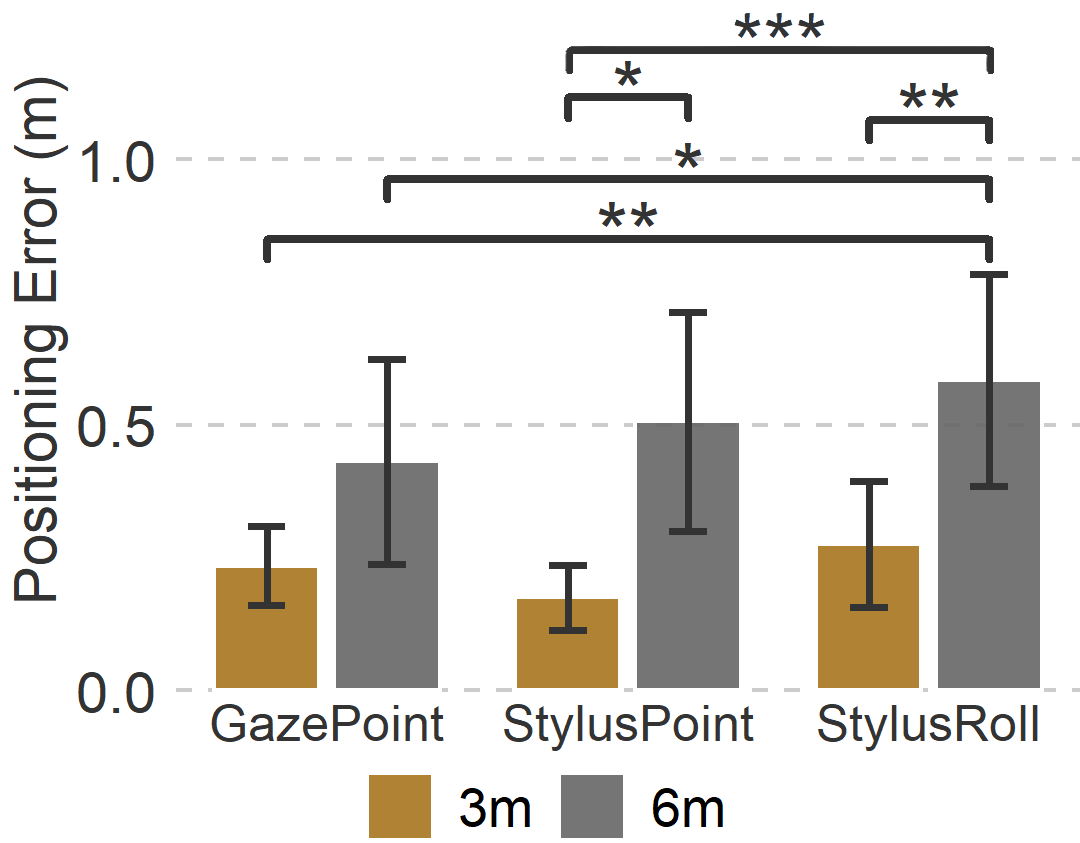}}
    \caption{Results of \PositioningError. \revised{Notably, main effects show \ModeFlip (a) and teleporting at 3m (b) to be more accurate.}}
    \label{fig:positioning-error}
\end{figure}

\revised{On \PositioningError (\autoref{fig:positioning-error}),} we found that participants were more accurate with \revised{\ModeFlip} (\revised{$F_{1, 187} = 28.86, p < 0.001, \eta^2_p=0.134$}).
\TargetDepthCondition also impacted \PositioningError, as the performance \revised{was} more accurate at the close depth of $3m$ (\revised{$F_{1, 187} = 25.87, p < 0.001, \eta^2_p=0.122$}).
\revised{Furthermore,} we found significant interactions between \ModeSwitchingCondition$\times$ \OrientationCondition (\revised{$F_{2, 187} = 4.33, p<0.015, \eta^2_p=0.044$}), \ModeSwitchingCondition $\times$ \TargetDepthCondition (\revised{$F_{1, 187} = 14.50, p<0.001, \eta^2_p=0.072$}), \OrientationCondition $\times$ \TargetDepthCondition (\revised{$F_{2, 187} = 4.015, p<0.020, \eta^2_p=0.041$}), and \ModeSwitchingCondition $\times$ \OrientationCondition $\times$ \TargetDepthCondition (\revised{$F_{2, 187} = 3.97, p<0.021, \eta^2_p=0.041$}). Specifically, \OrientationRoll\ \revised{was more impacted by \ModeButton} than both \OrientationPenPoint and \OrientationGazePoint (\revised{$p<0.041$}). Positioning \revised{error at $6m$} was \revised{also} significantly \revised{more impacted by \ModeButton} ($p<0.001$), \revised{while \OrientationRoll} was \revised{more impacted by \TargetDepthCondition} than \OrientationGazePoint ($p=0.017$). Finally, \revised{from the three-way interaction,} \OrientationGazePoint was more accurate \revised{than \OrientationRoll} \revised{when using} \ModeFlip\ \revised{while} teleporting $3m$ away ($p=0.019$).

\begin{figure}
    \centering
    \subfloat[\ModeSwitchingCondition]{\includegraphics[width=0.26\linewidth]{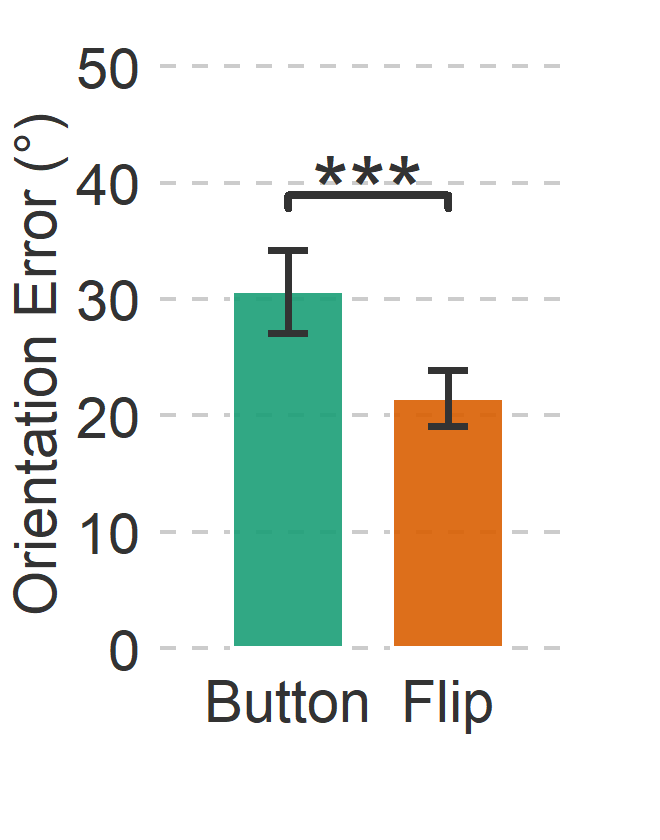}\label{subfig:ori-error-mode}}
    \hfill
    \subfloat[\TargetDepthCondition]{\includegraphics[width=0.26\linewidth]{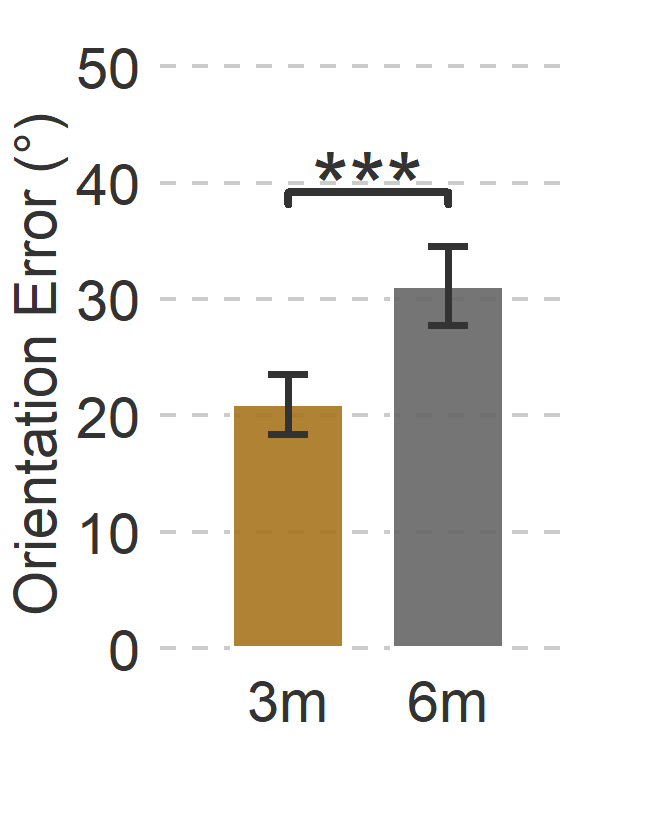}\label{subfig:ori-error-depth}}
    \hfill
    \subfloat[\ModeSwitchingCondition $\times$ \TargetDepthCondition.]{\includegraphics[width=0.41\linewidth]{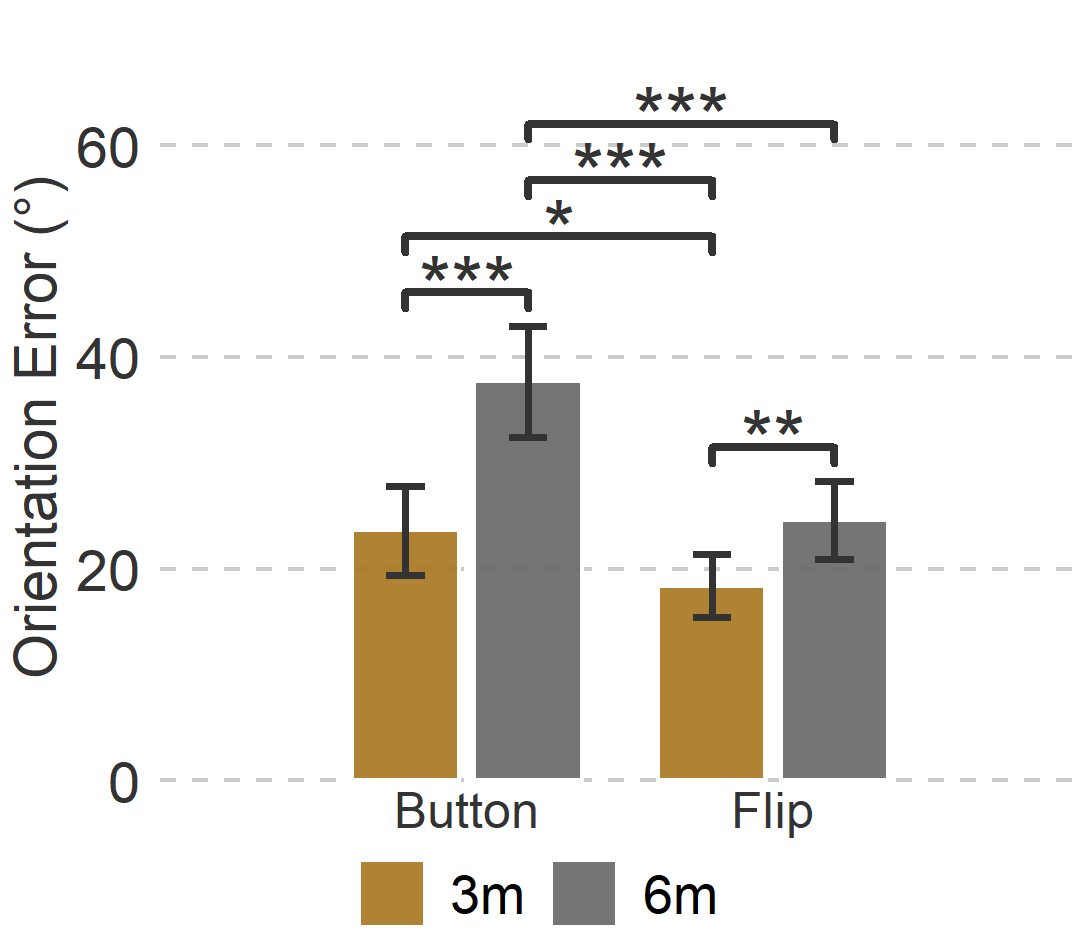}\label{subfig:ori-error-int-mode-depth}}
    \caption{Results of \OrientationError~\revised{(a-c). Notably, main effects show that \ModeFlip (a) and 3m (b) resulted in higher accuracy.}}
    \label{fig:orientation-error-and-preference}
\end{figure}

\revised{Regarding \OrientationError (\autoref{fig:orientation-error-and-preference}a–c),} we found that orientation was more accurate with \ModeFlip than \ModeButton (\revised{$F_{1, 187} = 28.02, p < 0.001, \eta^2_p=0.130$}) and at $3m$ compared with $6m$ (\revised{$F_{1, 187} = 54.54, p < 0.001, \eta^2_p=0.189$}). 
\revised{Additionally,} we found significant interactions between \ModeSwitchingCondition $\times$ \OrientationCondition (\revised{$F_{2, 187} = 3.22, p=0.042, \eta^2_p=0.033$}) and \ModeSwitchingCondition $\times$ \TargetDepthCondition (\revised{$F_{1, 187} = 6.25, p=0.013, \eta^2_p=0.032$}).
\revised{However, post hoc interaction analyses found no significant differences in the \ModeSwitchingCondition $\times$ \OrientationCondition interaction (all $p>0.078$).} \revised{Instead,} participants were more impacted by the difference in \TargetDepthCondition when using \ModeButton than when using \ModeFlip (\revised{$p<0.013$}).


\subsection{NASA-TLX, Preferences and User Feedback}
The only significant NASA-TLX result was that \OrientationGazePoint\!+\ModeButton was perceived less physically demanding ($\chi^2(5) = 15.625, p=0.008, W=0.935$) than \OrientationRoll\!+\ModeFlip ($p = 0.044$).

\begin{figure}
    \centering
    \includegraphics[width=0.55\linewidth]{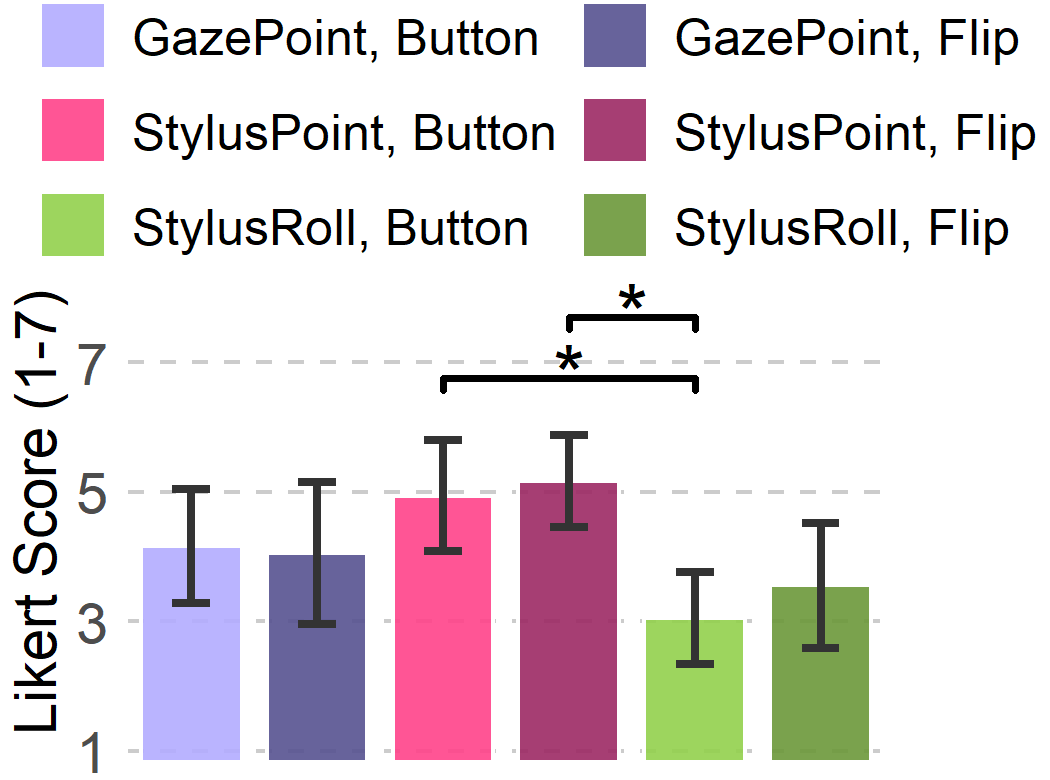}
    \caption{Preference results. ~\revised{Notably, main effects show that the \OrientationPenPoint variants are preferred by users over \ModeButton\!+\OrientationRoll}.}
    \label{fig:results-preference}
\end{figure}

The results on preference (\autoref{fig:results-preference}) indicated that participants  preferred \OrientationPenPoint (with both \ModeButton and \ModeFlip) over \OrientationRoll\!+\ModeButton (both $p < 0.037$). The feedback showed that they found all techniques viable for completing the task, but with trade-offs between \ModeSwitchingCondition and \OrientationCondition.

\subsubsection{\ModeButton vs. \ModeFlip}
The two \ModeSwitchingCondition methods divided opinions, reflecting a trade-off between intuitiveness and reliability. \ModeFlip was praised for its intuitiveness and reduced mental load since there was \textit{``no need to memorize [the] current mode''} (P3) and \textit{``I used less time [than] figuring out what the two buttons do''} (P5). However, some participants found \ModeFlip suffered from \textit{``awkward grips''} (P17) and made buttons \textit{``not very handy after reversing''} (P11). P7 even worried about the stylus \textit{``falling down''} when performing flips. In contrast, \ModeButton was seen as \textit{``more stable''} (P1, P15, P18) and provided \textit{``button feedback (click) so I know the mode has been switched''} (P7). 

\subsubsection{\OrientationRoll}
\OrientationRoll was perceived as \textit{``precise''} (P3, P11), but most participants (10 out of 18) criticized its physical demands. The most frequently mentioned problem was the strain and discomfort in their wrists, especially with \textit{``half circle (180$\degree$)''} rotations (P1, P8, P13), or with \textit{``clockwise rotations''} (P8). Although \OrientationRoll only maps roll-axis rotation to orientation, some participants still rotated the stylus freely along all axes to overcome the limited wrist rotational range. P9 noted that \textit{``I had to give up the precision for some large angles because my wrist was limited''}. In addition, holding the teleport button while rolling the stylus sometimes caused accidental releases (P9, P16, P17). 

\subsubsection{\OrientationPenPoint}
Controlling orientation by pointing with the stylus was considered the most \textit{``natural''}, \textit{``intuitive''}, and to provide \textit{``great control''} (P3, P6, P7, P9, P10, P12, P13, P15, P18). Participants reported that it was \textit{``much easier to hit the right rotation''} (P15). Compared with \OrientationRoll, participants found that \OrientationPenPoint caused less physical strain, as it was \textit{``much more efficient when the blackboard was reversed''} (P7, P16) and \textit{``became much easier to line up the orientation when being able to use all axes of the stylus''} (P15). 

\subsubsection{\OrientationGazePoint}
The gaze-based orientation method received mixed comments. Some praised it as \textit{``the fastest''} (P1, P3, P7), involving \textit{``less physical strain''} for the hand and being \textit{``effortless''} (P7, P13, P14, P16, P17). Others found it \textit{``fun''} (P4, P12, P13, P14), and \textit{``very intuitive''} (P3, P4, P13, P16, P17). However, several struggled with \textit{``eye-tracking [that] wasn't \revised{very accurate}''} (P5, P9, P10, P13, P16, P17).
Participants with glasses especially mentioned \textit{``eye fatigue''} (P8, P10, P12) and that \textit{``the eye tracker was pressing on my glasses uncomfortably''} (P17).

\subsection{Observations}
When switching modes via \ModeFlip, some participants 
\revised{reported struggling to change between the}
precision and palm grips. Though we anticipated a fluid transition that exploits finger dexterity, \ModeFlip was limited by \revised{individual differences} in hand size \revised{and motor ability across participants}. To prevent the stylus from slipping, some participants maintained the precision grip but rotated their wrist largely to point with the stylus tail, or even tried using their other hand to assist.


We also noted the hardware limitations of the Neon eye tracker. As a mounted component \revised{between the lenses and the user, it pressed} against participants' glasses and \revised{caused} discomfort. 
\revised{Though some participants who wore glasses reported eye tracking inaccuracy in \OrientationGazePoint, statistical analyses (ANOVA for normal, Kruskal-Wallis for non-normal) did not indicate significant differences in any performance metrics between the groups of participants with or without glasses. Therefore, we claim that despite discomfort, Neon XR provides reliable eye tracking quality for our study.} 



\section{Discussion}
In summary, we can address \textbf{RQ1}: \ModeFlip is \revised{faster} to perform and results in higher accuracy in subsequent teleportation. To address \textbf{RQ2}, we can conclude that \OrientationRoll is the most challenging for orientation control, 
while \OrientationPenPoint was most preferred by participants \revised{and was more efficient at orientation than \OrientationGazePoint}.

As guidelines for designing teleportation in stylus-based VR interfaces, we suggest assigning frequently-used functions to grip-based methods like \ModeFlip, which enables rapid switching, while binding less frequent functions to \ModeButton. Among orientation methods, \OrientationRoll is the least recommended, \revised{while \OrientationPenPoint becomes the most suitable choice for the system we tested.  \OrientationGazePoint achieves comparable task performance to \OrientationPenPoint, demonstrating its potential as an alternative orientation method. Potentially, a higher-accuracy eye-tracker can improve user performance with this method.}

In the following sections, we elaborate on these findings and discuss the study’s limitations and avenues for future work.

\subsection{Mode Switching Techniques}
\revised{A qualitative advantage of \ModeFlip over \ModeButton is that it does not rely on extra buttons. In contrast, the \ModeButton approach requires an extra mode-switch button, which may take away space for other operations. Yet, a button press is simple and efficient in principle compared with gestures.} 
\revised{Interestingly, our results show that switching between modes through the \ModeFlip was faster.
We attribute this} to intuitive action of flipping, as a natural behavior that is similar to pen use in physical contexts, compared with the arbitrary task of memorizing \revised{the current mode along with} locating \revised{and} pressing a specific button using a digital stylus, which requires dedicated visual feedback. 

\revised{Furthermore, while the act of positioning and orientation were slower using \ModeFlip, it achieved higher accuracy with comparable overall task time. }
The longer teleportation time occurs because with \ModeFlip, participants needed to switch grips and adapt to the palm grip--e.g., locating the teleport button--while \ModeButton maintained a static precision grip. \revised{The higher accuracy, however, corroborates} \citeauthor{Li20Grip}'s findings on stylus grips in VR: while pointing forward and downward is easier with a precision grip, pointing forward and upward is easier with a palm grip~\cite{Li20Grip}. Since parabola-based Point \& Teleport requires raising the stylus higher for distant targets, it becomes easier to use the palm grip after \ModeFlip than continuing with the precision grip in \ModeButton. 
\revised{This is further supported by that \ModeButton was more impacted by depth, with slower switching-out and lower orientation accuracy for far targets}
This indicates that the difficulty of holding the stylus higher with a precision grip reduces users' ability to finely control orientation.

\subsection{Orientation Control}
For the different orientation control methods, \revised{our} results show that participants took longer to complete \revised{the compound teleport-and-orient task} with \OrientationRoll than with the other two methods. \revised{This difference mainly occurs when orienting, where} \OrientationRoll was the slowest for orientation and \OrientationPenPoint was the fastest. We interpret these results as evidence that rolling the stylus to specify orientation is harder to perform because it requires fine hand dexterity to manipulate the stylus efficiently or effectively while holding it. 
\revised{This difficulty is augmented by the need to simultaneously press the teleport button, which restricts in-hand finger rotation and lead to more reliance on wrist rotation. } It is in line with other findings indicating its inherent difficulty \cite{Bozgeyikli16}. 

In contrast, pointing \revised{via the stylus or gaze} offers larger movement spaces and greater freedom, making orientation easier. Similarly, results for positioning errors indicate that \OrientationRoll produced worse accuracy when used together with \ModeButton, suggesting an additional challenge of rolling a stylus in a precision grip\revised{, which offers more limited rolling motion space than a palm grip}.

\revised{Regarding \OrientationGazePoint, although six participants reported eye-tracking inaccuracy, the orientation task itself does not require high-precision pointing, since users can physically fine-tune their orientation afterwards. Consequently, \OrientationGazePoint outperformed \OrientationRoll and showed no significant differences from \OrientationPenPoint in overall task time.
Looking one level deeper, we attribute this robustness to our design, which allows the orientation ray to intersect any surface rather than only the ground. By moving away from the strict assumption that teleport direction must be specified on the ground—and instead allowing orientation to be defined via arbitrary spatial objects—we leverage natural gaze behaviour: users tend to look toward the objects they want to inspect or interact with.}

\subsection{Limitations and Future Work}
Our explorations are grounded in the common Point \& Teleport approach, which defines teleportation position through hand (stylus) pointing. However, it is also possible to use gaze pointing for position specification \cite{Kim2023Exploration}, and it remains to be studied whether both position and orientation could be specified with gaze.
%
%
\revised{We conducted the study using a Meta Quest 3 with a Neon XR eye tracker, as it was the most appropriate hardware setup available for a stylus. It caused some discomfort and can be improved with better mounts or future devices that integrate eye-tracking and stylus. }
Some of our proposed techniques support setting orientation through pointing at object surfaces in 3D space, which diverges from most prior work that focused solely on ground-based teleportation, making it less comparable to prior approaches. Our study was conducted in a controlled laboratory setting, where participants could use only a limited set of surfaces for orientation, such as the blackboard, ground, and walls. Future work is needed to examine how these techniques translate to more realistic uses of a stylus involving more interactive objects and crowded environments. 
Also \revised{as \ModeFlip relies on stylus orientation, occasional unintentional flips can occur, e.g., when drawing on tables. Context-aware cues such as disabling \ModeFlip near surfaces have potential to address this.}
The study focused on two \revised{sub}tasks of positioning and orientation, though in many cases, users may only want to specify the position. It is plausible that the findings for the two mode-switching methods extend to \revised{purely} teleportation positioning, which could be validated by future work.
\revised{Further,} the study task required participants to specify a single teleport position and orientation per trial. While this sufficiently captures the fundamental teleportation task and likely extends to scenarios involving multiple successive teleports, further studies are needed to validate this assumption and generate additional insights.
%
%

\section{Conclusion}

\revised{In this paper,  we investigate teleportation for stylus-based interaction in VR through two mode-switching (\ModeButton and \ModeFlip gesture) and three orientation techniques (\OrientationRoll, \OrientationPenPoint, and \OrientationGazePoint). We evaluated these techniques in a compound point, orient, and teleport task. Our key takeaways are: (1) flipping the stylus to enter teleport mode is intuitive, makes users switch modes faster, and induces more accurate though slightly slower teleporting; (2) setting the landing orientation by pointing in the desired facing direction is easy to understand and more efficient than adjusting orientation via stylus roll. Pointing works with both a stylus-ray and gaze; these were on par in our tests, and more accurate eye-tracking could improve gaze-based orientation.

The interaction concepts can extend beyond styluses to controller- and bare-hand-based interaction. Controller teleportation already benefits from multiple buttons, but flip gestures and pointing for teleport orientation may offer additional flexibility. Bare-hand-based teleportation is underexplored, and opportunities may lie in leveraging flip-like gestural mode-switching and multimodal integration with gaze. This hints at a broader design space for advancing how users navigate virtual environments with intuitive multimodal inputs to explore in the future.}

\begin{acks}
    This work was supported by the Independent Research Fund Denmark (grant number 5254-00080B), a Google Research Gift award (``Multimodal and Gaze + Gesture Interactions in XR''), and the Danish National Research Foundation under the Pioneer Centre for AI in Denmark (DNRF grant P1).
\end{acks}

\bibliographystyle{ACM-Reference-Format}
\bibliography{refs}

@inproceedings{Arora2017Experimental,
    title        = {Experimental Evaluation of Sketching on Surfaces in VR},
    author       = {Arora, Rahul and Kazi, Rubaiat Habib and Anderson, Fraser and Grossman, Tovi and Singh, Karan and Fitzmaurice, George},
    year         = {2017},
    booktitle    = {Proceedings of the 2017 CHI Conference on Human Factors in Computing Systems},
    location     = {Denver, Colorado, USA},
    publisher    = {Association for Computing Machinery},
    address      = {New York, NY, USA},
    series       = {CHI '17},
    pages        = {5643–5654},
    doi          = {10.1145/3025453.3025474},
    isbn         = {9781450346559},
    url          = {https://doi.org/10.1145/3025453.3025474},
    numpages     = {12}
}

@inproceedings{Arora2018SymbiosisSketch,
    title        = {SymbiosisSketch: Combining 2D \& 3D Sketching for Designing Detailed 3D Objects in Situ},
    author       = {Arora, Rahul and Habib Kazi, Rubaiat and Grossman, Tovi and Fitzmaurice, George and Singh, Karan},
    year         = {2018},
    booktitle    = {Proceedings of the 2018 CHI Conference on Human Factors in Computing Systems},
    location     = {Montreal QC, Canada},
    publisher    = {Association for Computing Machinery},
    address      = {New York, NY, USA},
    series       = {CHI '18},
    pages        = {1–15},
    doi          = {10.1145/3173574.3173759},
    isbn         = {9781450356206},
    url          = {https://doi.org/10.1145/3173574.3173759},
    numpages     = {15}
}

@inproceedings{Batmaz20Precision,
    title        = {Precision vs. Power Grip: A Comparison of Pen Grip Styles for Selection in Virtual Reality},
    author       = {Batmaz, Anil Ufuk and Mutasim, Aunnoy K and Stuerzlinger, Wolfgang},
    year         = {2020},
    booktitle    = {2020 IEEE Conference on Virtual Reality and 3D User Interfaces Abstracts and Workshops (VRW)},
    pages        = {23--28},
    doi          = {10.1109/VRW50115.2020.00012}
}

@misc{Baumann23NeonTestReport,
    title        = {Neon Accuracy Test Report},
    author       = {Baumann, Chris and Dierkes, Kai},
    year         = {2023},
    month        = dec,
    publisher    = {Pupil Labs},
    doi          = {10.5281/zenodo.10420388},
    url          = {https://doi.org/10.5281/zenodo.10420388}
}

@inproceedings{Bi08,
    title        = {An exploration of pen rolling for pen-based interaction},
    author       = {Bi, Xiaojun and Moscovich, Tomer and Ramos, Gonzalo and Balakrishnan, Ravin and Hinckley, Ken},
    year         = {2008},
    booktitle    = {Proceedings of the 21st Annual ACM Symposium on User Interface Software and Technology},
    location     = {Monterey, CA, USA},
    publisher    = {Association for Computing Machinery},
    address      = {New York, NY, USA},
    series       = {UIST '08},
    pages        = {191–200},
    doi          = {10.1145/1449715.1449745},
    isbn         = {9781595939753},
    url          = {https://doi.org/10.1145/1449715.1449745},
    numpages     = {10}
}

@inproceedings{Bimberg2021Virtual,
    title        = {Virtual Rotations for Maneuvering in Immersive Virtual Environments},
    author       = {Bimberg, Pauline and Weissker, Tim and Kulik, Alexander and Froehlich, Bernd},
    year         = {2021},
    booktitle    = {Proceedings of the 27th ACM Symposium on Virtual Reality Software and Technology},
    publisher    = {Association for Computing Machinery},
    address      = {New York, NY, USA},
    series       = {VRST '21},
    pages        = {1–10},
    doi          = {10.1145/3489849.3489893},
    isbn         = {978-1-4503-9092-7},
    url          = {https://dl.acm.org/doi/10.1145/3489849.3489893},
    collection   = {VRST '21}
}

@inproceedings{Bozgeyikli16,
    title        = {Point \& Teleport Locomotion Technique for Virtual Reality},
    author       = {Bozgeyikli, Evren and Raij, Andrew and Katkoori, Srinivas and Dubey, Rajiv},
    year         = {2016},
    booktitle    = {Proceedings of the 2016 Annual Symposium on Computer-Human Interaction in Play},
    publisher    = {Association for Computing Machinery},
    address      = {New York, NY, USA},
    series       = {CHI PLAY '16},
    pages        = {205–216},
    doi          = {10.1145/2967934.2968105},
    isbn         = {978-1-4503-4456-2},
    url          = {https://dl.acm.org/doi/10.1145/2967934.2968105},
    collection   = {CHI PLAY '16}
}

@article{Bustamante2008MeasurementTLX,
    title        = {Measurement Invariance of the Nasa TLX},
    author       = {Bustamante, Ernesto A. and Spain, Randall D.},
    year         = {2008},
    month        = {9},
    journal      = {http://dx.doi.org/10.1177/154193120805201946},
    publisher    = {SAGE PublicationsSage CA: Los Angeles, CA},
    volume       = {3},
    pages        = {1522--1526},
    doi          = {10.1177/154193120805201946},
    isbn         = {9781605606859},
    issn         = {10711813},
    url          = {https://journals.sagepub.com/doi/10.1177/154193120805201946}
}

@article{Cai25HPIPainting,
    title        = {HPIPainting: A Hand-Pen Interaction for VR Painting},
    author       = {Cai, Ang and Yao, Chao and Liu, Hongjun and Li, Changsheng and Zhang, Yongyue and Guo, Yu and Wang, Xiaokun and Ban, Xiaojuan},
    year         = {2025},
    month        = sep,
    journal      = {Proc. ACM Interact. Mob. Wearable Ubiquitous Technol.},
    publisher    = {Association for Computing Machinery},
    address      = {New York, NY, USA},
    volume       = {9},
    number       = {3},
    doi          = {10.1145/3749538},
    url          = {https://doi.org/10.1145/3749538},
    issue_date   = {September 2025},
    articleno    = {71},
    numpages     = {26}
}

@inproceedings{Cami18,
    title        = {Unimanual Pen+Touch Input Using Variations of Precision Grip Postures},
    author       = {Cami, Drini and Matulic, Fabrice and Calland, Richard G. and Vogel, Brian and Vogel, Daniel},
    year         = {2018},
    booktitle    = {Proceedings of the 31st Annual ACM Symposium on User Interface Software and Technology},
    location     = {Berlin, Germany},
    publisher    = {Association for Computing Machinery},
    address      = {New York, NY, USA},
    series       = {UIST '18},
    pages        = {825–837},
    doi          = {10.1145/3242587.3242652},
    isbn         = {9781450359481},
    url          = {https://doi.org/10.1145/3242587.3242652},
    numpages     = {13}
}

@inproceedings{Chen22Investigating,
    title        = {Investigating Input Modality and Task Geometry on Precision-first 3D Drawing in Virtual Reality},
    author       = {Chen, Chen and Yarmand, Matin and Xu, Zhuoqun and Singh, Varun and Zhang, Yang and Weibel, Nadir},
    year         = {2022},
    booktitle    = {2022 IEEE International Symposium on Mixed and Augmented Reality (ISMAR)},
    pages        = {384--393},
    doi          = {10.1109/ISMAR55827.2022.00054}
}

@article{Cherep2020Spatial,
    title        = {Spatial cognitive implications of teleporting through virtual environments},
    author       = {Cherep, Lucia A. and Lim, Alex F. and Kelly, Jonathan W. and Acharya, Devi and Velasco, Alfredo and Bustamante, Emanuel and Ostrander, Alec G. and Gilbert, Stephen B.},
    year         = {2020},
    journal      = {Journal of Experimental Psychology: Applied},
    publisher    = {American Psychological Association},
    address      = {US},
    volume       = {26},
    number       = {3},
    pages        = {480–492},
    doi          = {10.1037/xap0000263},
    issn         = {1939-2192},
    abstractnote = {Teleporting is a popular interface to allow virtual reality users to explore environments that are larger than the available walking space. When teleporting, the user positions a marker in the virtual environment and is instantly transported without any self-motion cues. Five experiments were designed to evaluate the spatial cognitive consequences of teleporting and to identify environmental cues that could mitigate those costs. Participants performed a triangle completion task by traversing 2 outbound path legs before pointing to the unmarked path origin. Locomotion was accomplished via walking or 2 common implementations of the teleporting interface distinguished by the concordance between movement of the body and movement through the virtual environment. In the partially concordant teleporting interface, participants teleported to translate (change position) but turned the body to rotate. In the discordant teleporting interface, participants teleported to translate and rotate. Across all 5 experiments, discordant teleporting produced larger errors than partially concordant teleporting which produced larger errors than walking, reflecting the importance of translational and rotational self-motion cues. Furthermore, geometric boundaries (room walls or a fence) were necessary to mitigate the spatial cognitive costs associated with teleporting, and landmarks were helpful only in the context of a geometric boundary. (PsycInfo Database Record (c) 2020 APA, all rights reserved)}
}

@article{Deering1995HoloSketch,
    title        = {HoloSketch: a virtual reality sketching/animation tool},
    author       = {Deering, Michael F.},
    year         = {1995},
    month        = sep,
    journal      = {ACM Trans. Comput.-Hum. Interact.},
    publisher    = {Association for Computing Machinery},
    address      = {New York, NY, USA},
    volume       = {2},
    number       = {3},
    pages        = {220–238},
    doi          = {10.1145/210079.210087},
    issn         = {1073-0516},
    url          = {https://doi.org/10.1145/210079.210087},
    issue_date   = {Sept. 1995},
    numpages     = {19}
}

@inproceedings{Drey20VRSketchIn,
    title        = {VRSketchIn: Exploring the Design Space of Pen and Tablet Interaction for 3D Sketching in Virtual Reality},
    author       = {Drey, Tobias and Gugenheimer, Jan and Karlbauer, Julian and Milo, Maximilian and Rukzio, Enrico},
    year         = {2020},
    booktitle    = {Proceedings of the 2020 CHI Conference on Human Factors in Computing Systems},
    location     = {Honolulu, HI, USA},
    publisher    = {Association for Computing Machinery},
    address      = {New York, NY, USA},
    series       = {CHI '20},
    pages        = {1–14},
    doi          = {10.1145/3313831.3376628},
    isbn         = {9781450367080},
    url          = {https://doi.org/10.1145/3313831.3376628},
    numpages     = {14}
}

@misc{ENGAGE,
    title        = {Spatial Computing Artificial Intelligence - ENGAGE XR},
    author       = {ENGAGE},
    year         = {2025},
    url          = {https://engagevr.io/}
}

@inproceedings{Funk19,
    title        = {Assessing the Accuracy of Point \& Teleport Locomotion with Orientation Indication for Virtual Reality using Curved Trajectories},
    author       = {Funk, Markus and M\"{u}ller, Florian and Fendrich, Marco and Shene, Megan and Kolvenbach, Moritz and Dobbertin, Niclas and G\"{u}nther, Sebastian and M\"{u}hlh\"{a}user, Max},
    year         = {2019},
    month        = may,
    booktitle    = {Proceedings of the 2019 CHI Conference on Human Factors in Computing Systems},
    publisher    = {ACM},
    address      = {Glasgow Scotland Uk},
    pages        = {1–12},
    doi          = {10.1145/3290605.3300377},
    isbn         = {978-1-4503-5970-2},
    url          = {https://dl.acm.org/doi/10.1145/3290605.3300377},
    language     = {en}
}

@misc{GravitySketch,
    title        = {Gravity Sketch},
    author       = {Gravity Sketch},
    year         = {2025},
    url          = {https://gravitysketch.com/}
}

@inproceedings{Hachet2008Navidget,
    title        = {Navidget for Easy 3D Camera Positioning from 2D Inputs},
    author       = {Hachet, Martin and Decle, Fabrice and Knodel, Sebastian and Guitton, Pascal},
    year         = {2008},
    booktitle    = {2008 IEEE Symposium on 3D User Interfaces},
    pages        = {83--89},
    doi          = {10.1109/3DUI.2008.4476596}
}

@article{Hart1988DevelopmentResearch,
    title        = {Development of NASA-TLX (Task Load Index): Results of Empirical and Theoretical Research},
    author       = {Hart, Sandra G. and Staveland, Lowell E.},
    year         = {1988},
    month        = {1},
    journal      = {Advances in Psychology},
    publisher    = {North-Holland},
    volume       = {52},
    number       = {C},
    pages        = {139--183},
    doi          = {10.1016/S0166-4115(08)62386-9},
    issn         = {0166-4115}
}

@article{Hart2006Nasa-TaskLater,
    title        = {Nasa-Task Load Index (NASA-TLX); 20 Years Later},
    author       = {Hart, Sandra G.},
    year         = {2006},
    month        = {10},
    journal      = {http://dx.doi.org/10.1177/154193120605000909},
    publisher    = {SAGE PublicationsSage CA: Los Angeles, CA},
    pages        = {904--908},
    doi          = {10.1177/154193120605000909},
    isbn         = {9780945289296},
    issn         = {10711813},
    url          = {https://journals.sagepub.com/doi/abs/10.1177/154193120605000909}
}

@article{Jacob16,
    title        = {What you look at is what you get: gaze-based user interfaces},
    author       = {Jacob, Rob and Stellmach, Sophie},
    year         = {2016},
    month        = aug,
    journal      = {Interactions},
    publisher    = {Association for Computing Machinery},
    address      = {New York, NY, USA},
    volume       = {23},
    number       = {5},
    pages        = {62–65},
    doi          = {10.1145/2978577},
    issn         = {1072-5520},
    url          = {https://doi.org/10.1145/2978577},
    issue_date   = {September + October 2016},
    numpages     = {4}
}

@inproceedings{Jacob90,
    title        = {What You Look at is What You Get: Eye Movement-Based Interaction Techniques},
    author       = {Jacob, Robert J. K.},
    year         = {1990},
    booktitle    = {Proceedings of the SIGCHI Conference on Human Factors in Computing Systems},
    location     = {Seattle, Washington, USA},
    publisher    = {Association for Computing Machinery},
    address      = {New York, NY, USA},
    series       = {CHI '90},
    pages        = {11–18},
    doi          = {10.1145/97243.97246},
    isbn         = {0201509326},
    url          = {https://doi.org/10.1145/97243.97246},
    numpages     = {8}
}

@inproceedings{Kang2024RayHand,
    title        = {The RayHand Navigation: A Virtual Navigation Method with Relative Position between Hand and Gaze-Ray},
    author       = {Kang, Sei and Jeong, Jaejoon and Lee, Gun A. and Kim, Soo-Hyung and Yang, Hyung-Jeong and Kim, Seungwon},
    year         = {2024},
    booktitle    = {Proceedings of the 2024 CHI Conference on Human Factors in Computing Systems},
    location     = {Honolulu, HI, USA},
    publisher    = {Association for Computing Machinery},
    address      = {New York, NY, USA},
    series       = {CHI '24},
    doi          = {10.1145/3613904.3642147},
    isbn         = {9798400703300},
    url          = {https://doi.org/10.1145/3613904.3642147},
    articleno    = {634},
    numpages     = {15}
}

@article{Keefe2007drawing,
    title        = {Drawing on Air: Input Techniques for Controlled 3D Line Illustration},
    author       = {Keefe, Daniel and Zeleznik, Robert and Laidlaw, David},
    year         = {2007},
    journal      = {IEEE Transactions on Visualization and Computer Graphics},
    volume       = {13},
    number       = {5},
    pages        = {1067--1081},
    doi          = {10.1109/TVCG.2007.1060}
}

@inproceedings{Kehoe2025Logitech,
    title        = {Logitech MX Ink: Extending the Meta Quest Platform with a 6DoF Stylus},
    author       = {Kehoe, Aidan and Gutierrez, Mario and Ahn, Yena and Kogan, Vadim},
    year         = {2025},
    booktitle    = {2025 IEEE Conference on Virtual Reality and 3D User Interfaces Abstracts and Workshops (VRW)},
    pages        = {338--342},
    doi          = {10.1109/VRW66409.2025.00080}
}

@article{Kim2023Exploration,
    title        = {Exploration of the Virtual Reality Teleportation Methods Using Hand-Tracking, Eye-Tracking, and EEG},
    author       = {Kim, Jinwook and Jang, Hyunyoung and Kim, Dooyoung and Lee, Jeongmi},
    year         = {2023},
    month        = dec,
    journal      = {International Journal of Human–Computer Interaction},
    publisher    = {Taylor \& Francis},
    volume       = {39},
    number       = {20},
    pages        = {4112–4125},
    doi          = {10.1080/10447318.2022.2109248},
    issn         = {1044-7318}
}

@inproceedings{Knödel2008Navidget,
    title        = {Navidget for immersive virtual environments},
    author       = {Kn\"{o}del, Sebastian and Hachet, Martin and Guitton, Pascal},
    year         = {2008},
    booktitle    = {Proceedings of the 2008 ACM symposium on Virtual reality software and technology},
    publisher    = {Association for Computing Machinery},
    address      = {New York, NY, USA},
    series       = {VRST '08},
    pages        = {47–50},
    doi          = {10.1145/1450579.1450589},
    isbn         = {978-1-59593-951-7},
    url          = {https://dl.acm.org/doi/10.1145/1450579.1450589},
    collection   = {VRST '08}
}

@article{Lee2024RPG,
    title        = {RPG: Rotation Technique in VR Locomotion using Peripheral Gaze},
    author       = {Lee, Jaeyoon and Kim, Hanseob and Yang, Yechan and Kim, Gerard Jounghyun},
    year         = {2024},
    month        = may,
    journal      = {Proc. ACM Hum.-Comput. Interact.},
    publisher    = {Association for Computing Machinery},
    address      = {New York, NY, USA},
    volume       = {8},
    number       = {ETRA},
    doi          = {10.1145/3655609},
    url          = {https://doi.org/10.1145/3655609},
    issue_date   = {May 2024},
    articleno    = {235},
    numpages     = {19}
}

@inproceedings{Lee2024Snap,
    title        = {Snap, Pursuit and Gain: Virtual Reality Viewport Control by Gaze},
    author       = {Lee, Hock Siang and Weidner, Florian and Sidenmark, Ludwig and Gellersen, Hans},
    year         = {2024},
    booktitle    = {Proceedings of the 2024 CHI Conference on Human Factors in Computing Systems},
    location     = {Honolulu, HI, USA},
    publisher    = {Association for Computing Machinery},
    address      = {New York, NY, USA},
    series       = {CHI '24},
    doi          = {10.1145/3613904.3642838},
    isbn         = {9798400703300},
    url          = {https://doi.org/10.1145/3613904.3642838},
    articleno    = {375},
    numpages     = {14}
}

@inproceedings{Li2005Mode,
    title        = {Experimental analysis of mode switching techniques in pen-based user interfaces},
    author       = {Li, Yang and Hinckley, Ken and Guan, Zhiwei and Landay, James A.},
    year         = {2005},
    booktitle    = {Proceedings of the SIGCHI Conference on Human Factors in Computing Systems},
    location     = {Portland, Oregon, USA},
    publisher    = {Association for Computing Machinery},
    address      = {New York, NY, USA},
    series       = {CHI '05},
    pages        = {461–470},
    doi          = {10.1145/1054972.1055036},
    isbn         = {1581139985},
    url          = {https://doi.org/10.1145/1054972.1055036},
    numpages     = {10}
}

@inproceedings{Li20Grip,
    title        = {Get a Grip: Evaluating Grip Gestures for VR Input using a Lightweight Pen},
    author       = {Li, Nianlong and Han, Teng and Tian, Feng and Huang, Jin and Sun, Minghui and Irani, Pourang and Alexander, Jason},
    year         = {2020},
    booktitle    = {Proceedings of the 2020 CHI Conference on Human Factors in Computing Systems},
    location     = {Honolulu, HI, USA},
    publisher    = {Association for Computing Machinery},
    address      = {New York, NY, USA},
    series       = {CHI '20},
    pages        = {1–13},
    doi          = {10.1145/3313831.3376698},
    isbn         = {9781450367080},
    url          = {https://doi.org/10.1145/3313831.3376698},
    numpages     = {13}
}

@inproceedings{Lystbaek24,
    title        = {Hands-on, Hands-off: Gaze-Assisted Bimanual 3D Interaction},
    author       = {Lystb\ae{}k, Mathias N. and Mikkelsen, Thorbj\o{}rn and Krisztandl, Roland and Gonzalez, Eric J and Gonzalez-Franco, Mar and Gellersen, Hans and Pfeuffer, Ken},
    year         = {2024},
    booktitle    = {Proceedings of the 37th Annual ACM Symposium on User Interface Software and Technology},
    location     = {Pittsburgh, PA, USA},
    publisher    = {Association for Computing Machinery},
    address      = {New York, NY, USA},
    series       = {UIST '24},
    doi          = {10.1145/3654777.3676331},
    isbn         = {9798400706288},
    url          = {https://doi.org/10.1145/3654777.3676331},
    articleno    = {80},
    numpages     = {12}
}

@inproceedings{Matviienko22,
    title        = {SkyPort: Investigating 3D Teleportation Methods in Virtual Environments},
    author       = {Matviienko, Andrii and M\"{u}ller, Florian and Schmitz, Martin and Fendrich, Marco and M\"{u}hlh\"{a}user, Max},
    year         = {2022},
    booktitle    = {Proceedings of the 2022 CHI Conference on Human Factors in Computing Systems},
    publisher    = {Association for Computing Machinery},
    address      = {New York, NY, USA},
    series       = {CHI '22},
    pages        = {1–11},
    doi          = {10.1145/3491102.3501983},
    isbn         = {978-1-4503-9157-3},
    url          = {https://dl.acm.org/doi/10.1145/3491102.3501983},
    collection   = {CHI '22}
}

@misc{Meta25NLocomotion,
    title        = {Locomotion types},
    author       = {Meta Horizon},
    year         = {2025},
    publisher    = {Meta Horizon},
    url          = {https://developers.meta.com/horizon/design/locomotion-types/}
}

@article{mine95,
    title        = {Virtual environment interaction techniques},
    author       = {Mine, Mark R},
    year         = {1995},
    journal      = {UNC Chapel Hill CS Dept}
}

@article{mori2023point,
    title        = {Point \& Teleport with Orientation Specification, Revisited: Is Natural Turning Always Superior?},
    author       = {Mori, Shohei and Hashiguchi, Satoshi and Shibata, Fumihisa and Kimura, Asako},
    year         = {2023},
    journal      = {Journal of Information Processing},
    publisher    = {Information Processing Society of Japan},
    volume       = {31},
    pages        = {392--403}
}

@inproceedings{Muller2023UndoPort,
    title        = {UndoPort: Exploring the Influence of Undo-Actions for Locomotion in Virtual Reality on the Efficiency, Spatial Understanding and User Experience},
    author       = {M\"{u}ller, Florian and Ye, Arantxa and Sch\"{o}n, Dominik and Rasch, Julian},
    year         = {2023},
    booktitle    = {Proceedings of the 2023 CHI Conference on Human Factors in Computing Systems},
    location     = {Hamburg, Germany},
    publisher    = {Association for Computing Machinery},
    address      = {New York, NY, USA},
    series       = {CHI '23},
    doi          = {10.1145/3544548.3581557},
    isbn         = {9781450394215},
    url          = {https://doi.org/10.1145/3544548.3581557},
    articleno    = {234},
    numpages     = {15}
}

@inproceedings{Pfeuffer17pinch,
    title        = {Gaze + pinch interaction in virtual reality},
    author       = {Pfeuffer, Ken and Mayer, Benedikt and Mardanbegi, Diako and Gellersen, Hans},
    year         = {2017},
    booktitle    = {Proceedings of the 5th Symposium on Spatial User Interaction},
    location     = {Brighton, United Kingdom},
    publisher    = {Association for Computing Machinery},
    address      = {New York, NY, USA},
    series       = {SUI '17},
    pages        = {99–108},
    doi          = {10.1145/3131277.3132180},
    isbn         = {9781450354868},
    url          = {https://doi.org/10.1145/3131277.3132180},
    numpages     = {10}
}

@inproceedings{Pfeuffer20,
    title        = {Empirical Evaluation of Gaze-Enhanced Menus in Virtual Reality},
    author       = {Pfeuffer, Ken and Mecke, Lukas and Delgado Rodriguez, Sarah and Hassib, Mariam and Maier, Hannah and Alt, Florian},
    year         = {2020},
    booktitle    = {26th ACM Symposium on Virtual Reality Software and Technology},
    location     = {Virtual Event, Canada},
    publisher    = {Association for Computing Machinery},
    address      = {New York, NY, USA},
    series       = {VRST '20},
    doi          = {10.1145/3385956.3418962},
    isbn         = {9781450376198},
    url          = {https://doi.org/10.1145/3385956.3418962},
    articleno    = {20},
    numpages     = {11}
}

@inproceedings{Pfeuffer2014touch,
    title        = {Gaze-touch: combining gaze with multi-touch for interaction on the same surface},
    author       = {Pfeuffer, Ken and Alexander, Jason and Chong, Ming Ki and Gellersen, Hans},
    year         = {2014},
    booktitle    = {Proceedings of the 27th Annual ACM Symposium on User Interface Software and Technology},
    location     = {Honolulu, Hawaii, USA},
    publisher    = {Association for Computing Machinery},
    address      = {New York, NY, USA},
    series       = {UIST '14},
    pages        = {509–518},
    doi          = {10.1145/2642918.2647397},
    isbn         = {9781450330695},
    url          = {https://doi.org/10.1145/2642918.2647397},
    numpages     = {10}
}

@inproceedings{Pfeuffer2015Shift,
    title        = {Gaze-Shifting: Direct-Indirect Input with Pen and Touch Modulated by Gaze},
    author       = {Pfeuffer, Ken and Alexander, Jason and Chong, Ming Ki and Zhang, Yanxia and Gellersen, Hans},
    year         = {2015},
    booktitle    = {Proceedings of the 28th Annual ACM Symposium on User Interface Software \& Technology},
    location     = {Charlotte, NC, USA},
    publisher    = {Association for Computing Machinery},
    address      = {New York, NY, USA},
    series       = {UIST '15},
    pages        = {373–383},
    doi          = {10.1145/2807442.2807460},
    isbn         = {9781450337793},
    url          = {https://doi.org/10.1145/2807442.2807460},
    numpages     = {11}
}

@inproceedings{Qiu25,
    title        = {MaRginalia: Enabling In-person Lecture Capturing and Note-taking Through Mixed Reality},
    author       = {Qiu, Leping and Kim, Erin Seongyoon and Suh, Sangho and Sidenmark, Ludwig and Grossman, Tovi},
    year         = {2025},
    booktitle    = {Proceedings of the 2025 CHI Conference on Human Factors in Computing Systems},
    location     = {},
    publisher    = {Association for Computing Machinery},
    address      = {New York, NY, USA},
    series       = {CHI '25},
    doi          = {10.1145/3706598.3714065},
    isbn         = {9798400713941},
    url          = {https://doi.org/10.1145/3706598.3714065},
    articleno    = {141},
    numpages     = {15}
}

@inproceedings{Romat21Flashpen,
    title        = {Flashpen: A High-Fidelity and High-Precision Multi-Surface Pen for Virtual Reality},
    author       = {Romat, Hugo and Fender, Andreas and Meier, Manuel and Holz, Christian},
    year         = {2021},
    booktitle    = {2021 IEEE Virtual Reality and 3D User Interfaces (VR)},
    pages        = {306--315},
    doi          = {10.1109/VR50410.2021.00053}
}

@inproceedings{Saund03Stylus,
    title        = {Stylus input and editing without prior selection of mode},
    author       = {Saund, Eric and Lank, Edward},
    year         = {2003},
    booktitle    = {Proceedings of the 16th Annual ACM Symposium on User Interface Software and Technology},
    location     = {Vancouver, Canada},
    publisher    = {Association for Computing Machinery},
    address      = {New York, NY, USA},
    series       = {UIST '03},
    pages        = {213–216},
    doi          = {10.1145/964696.964720},
    isbn         = {1581136366},
    url          = {https://doi.org/10.1145/964696.964720},
    numpages     = {4}
}

@misc{ShapesXR,
    title        = {Bring ideas to life in 3D},
    author       = {ShapesXR},
    year         = {2025},
    url          = {https://www.shapesxr.com/}
}

@inproceedings{Smith20Evaluating,
    title        = {Evaluating the Scalability of Non-Preferred Hand Mode Switching in Augmented Reality},
    author       = {Smith, Jesse and Wang, Isaac and Wei, Winston and Woodward, Julia and Ruiz, Jaime},
    year         = {2020},
    booktitle    = {Proceedings of the 2020 International Conference on Advanced Visual Interfaces},
    location     = {Salerno, Italy},
    publisher    = {Association for Computing Machinery},
    address      = {New York, NY, USA},
    series       = {AVI '20},
    doi          = {10.1145/3399715.3399850},
    isbn         = {9781450375351},
    url          = {https://doi.org/10.1145/3399715.3399850},
    articleno    = {19},
    numpages     = {9}
}

@inproceedings{Song_2011_Grips,
    title        = {Grips and gestures on a multi-touch pen},
    author       = {Song, Hyunyoung and Benko, Hrvoje and Guimbretiere, Francois and Izadi, Shahram and Cao, Xiang and Hinckley, Ken},
    year         = {2011},
    booktitle    = {Proceedings of the SIGCHI Conference on Human Factors in Computing Systems},
    location     = {Vancouver, BC, Canada},
    publisher    = {Association for Computing Machinery},
    address      = {New York, NY, USA},
    series       = {CHI '11},
    pages        = {1323–1332},
    doi          = {10.1145/1978942.1979138},
    isbn         = {9781450302289},
    url          = {https://doi.org/10.1145/1978942.1979138},
    numpages     = {10}
}

@inproceedings{Surale2019Mode,
    title        = {Experimental Analysis of Barehand Mid-air Mode-Switching Techniques in Virtual Reality},
    author       = {Surale, Hemant Bhaskar and Matulic, Fabrice and Vogel, Daniel},
    year         = {2019},
    booktitle    = {Proceedings of the 2019 CHI Conference on Human Factors in Computing Systems},
    location     = {Glasgow, Scotland Uk},
    publisher    = {Association for Computing Machinery},
    address      = {New York, NY, USA},
    series       = {CHI '19},
    pages        = {1–14},
    doi          = {10.1145/3290605.3300426},
    isbn         = {9781450359702},
    url          = {https://doi.org/10.1145/3290605.3300426},
    numpages     = {14}
}

@article{Templeman99,
    title        = {Virtual Locomotion: Walking in Place through Virtual Environments},
    author       = {Templeman, James N. and Denbrook, Patricia S. and Sibert, Linda E.},
    year         = {1999},
    month        = dec,
    journal      = {Presence: Teleoper. Virtual Environ.},
    publisher    = {MIT Press},
    address      = {Cambridge, MA, USA},
    volume       = {8},
    number       = {6},
    pages        = {598–617},
    doi          = {10.1162/105474699566512},
    issn         = {1054-7460},
    url          = {https://doi.org/10.1162/105474699566512},
    issue_date   = {December 1999},
    numpages     = {20}
}

@misc{VictoryXR,
    title        = {VictoryXR},
    author       = {VictoryXR},
    year         = {2025},
    url          = {https://www.victoryxr.com/}
}

@article{vogel10,
    title        = {Direct pen interaction with a conventional graphical user interface},
    author       = {Vogel, Daniel and Balakrishnan, Ravin},
    year         = {2010},
    journal      = {Human--Computer Interaction},
    publisher    = {Taylor \& Francis},
    volume       = {25},
    number       = {4},
    pages        = {324--388}
}

@inproceedings{wagner2024eye,
    title        = {Eye-Hand Movement of Objects in Near Space Extended Reality},
    author       = {Wagner, Uta and Asferg Jacobsen, Andreas and Feuchtner, Tiare and Gellersen, Hans and Pfeuffer, Ken},
    year         = {2024},
    booktitle    = {Proceedings of the 37th Annual ACM Symposium on User Interface Software and Technology},
    pages        = {1--13}
}

@incollection{wagner25pen,
    title        = {A Study of Multimodal Pen+ Gaze Interaction Techniques for Shape Point Translation in Extended Reality},
    author       = {Wagner, Uta and Kim, Jinwook and Wu, Zhikun and Zhou, Qiushi and Romero, Mario and Iop, Alessandro and Feuchtner, Tiare and Pfeuffer, Ken},
    year         = {2025},
    month        = {08},
    booktitle    = {Proceedings - 2025 IEEE International Symposium on Mixed and Augmented Reality, ISMAR 2025},
    doi = {10.1109/ISMAR67309.2025.00056}
}

@article{Weissker2023Gaining,
    title        = {Gaining the High Ground: Teleportation to Mid-Air Targets in Immersive Virtual Environments},
    author       = {Weissker, Tim and Bimberg, Pauline and Gokhale, Aalok Shashidhar and Kuhlen, Torsten and Froehlich, Bernd},
    year         = {2023},
    month        = may,
    journal      = {IEEE Transactions on Visualization and Computer Graphics},
    volume       = {29},
    number       = {5},
    pages        = {2467–2477},
    doi          = {10.1109/TVCG.2023.3247114},
    issn         = {1941-0506}
}

@ARTICLE{Rupp2025HowFar,
  author={Rupp, Daniel and Weissker, Tim and Wölwer, Matthias and Kuhlen, Torsten W. and Zielasko, Daniel},
  journal={IEEE Transactions on Visualization and Computer Graphics}, 
  title={How Far is Too Far? the Trade-Off Between Selection Distance and Accuracy During Teleportation in Immersive Virtual Reality}, 
  year={2025},
  volume={},
  number={},
  pages={1-15},
  doi={10.1109/TVCG.2025.3632345}
}

@inproceedings{Matulic2023PenTouchMidairCH,
  title={Pen+Touch+Midair: Cross-Space Hybrid Bimanual Interaction on Horizontal Surfaces in Virtual Reality},
  author={Fabrice Matulic and Daniel Vogel},
  year={2023},
  url={https://api.semanticscholar.org/CorpusID:259300598}
}

@article{Sermarini21Kinematic,
author = {John Sermarini and Joseph T. KiderJr. and Joseph J. LaViolaJr. and Daniel S. McConnell},
title ={A Kinematic Evaluation of Linear and Parabolic Pointing in Virtual Reality},

journal = {Proceedings of the Human Factors and Ergonomics Society Annual Meeting},
volume = {65},
number = {1},
pages = {86-90},
year = {2021},
doi = {10.1177/1071181321651131},
}

@INPROCEEDINGS{Narbayev25,
  author={Narbayev, Bakdauren and Ullah, A K M Amanat and Sin, Jaisie and Lasserre, Patricia and Hasan, Khalad},
  booktitle={2025 IEEE International Symposium on Mixed and Augmented Reality (ISMAR)}, 
  title={Exploring Pointing and Confirmation Techniques for Teleportation Across Varying Elevations in Virtual Reality}, 
  year={2025},
  volume={},
  number={},
  pages={1671-1681},
  doi={10.1109/ISMAR67309.2025.00170}
}

@misc{Arkio,
    title        = {Experience the unbuilt, together},
    author       = {Arkio},
    year         = {2025},
    url          = {https://www.arkio.is/}
}

\appendix

\end{document}